\documentclass[%
 reprint,
nofootinbib,
 amsmath,amssymb,
 aps,
]{revtex4-2}

\usepackage{graphicx}
\usepackage{dcolumn}
\usepackage{bm}
\usepackage{xcolor}
\usepackage[normalem]{ulem} 
\usepackage[hidelinks]{hyperref}

\newcommand{\mbf}[1]{\mathbf{#1}}
\newcommand{\matr}[1]{\mathbf{#1}}
\newcommand{\gmatr}[1]{\bm{#1}}
\newcommand{\vect}[1]{\bm{#1}}
\DeclareMathAlphabet\mathbfcal{OMS}{cmsy}{b}{n}
\DeclareMathOperator{\dprime}{\prime \prime}

\newcommand{\comment}[1]{{\color{blue}#1}}

\begin{document}

\preprint{APS/123-QED}

\title{Removing systematics-induced 21-cm foreground residuals by cross-correlating filtered data}

\author{Haochen~Wang$^{1,2}$}
\email[]{hcwang96@mit.edu}
\author{Juan~Mena-Parra$^{2}$}
\author{Tianyue~Chen$^{1,2,3}$}
\author{Kiyoshi~Masui$^{1,2}$}
\affiliation{$^1$ Department of Physics, Massachusetts Institute of Technology, 77 Massachusetts Avenue Cambridge, MA 02139, USA}
\affiliation{$^2$ MIT Kavli Institute for Astrophysics and Space Research, Massachusetts Institute of Technology, 77 Massachusetts Avenue Cambridge, MA 02139, USA}
\affiliation{$^3$ Institute of Physics, Laboratory of Astrophysics, Ecole Polytechnique F\'{e}d\'{e}rale de Lausanne (EPFL), Observatoire de Sauverny, 1290 Versoix, Switzerland}

\date{\today}

\begin{abstract}
Observations of the redshifted 21-cm signal 
emitted by neutral hydrogen represent a promising 
probe of large-scale structure in the universe. However, cosmological 
21-cm signal is challenging to observe due to astrophysical 
foregrounds which are several orders of magnitude brighter.
Traditional linear foreground removal methods can optimally remove 
foregrounds for a known telescope response but are sensitive to telescope 
systematic errors such as antenna gain and delay errors, leaving foreground 
contamination in the recovered signal. Non-linear 
methods such as principal component analysis, on the other hand, 
have been used successfully for foreground removal, but they lead to
signal loss that is difficult to characterize and requires careful
analysis. In this paper, we present a systematics-robust foreground 
removal technique which combines both linear and non-linear methods. 
We first obtain signal and foreground estimates using a 
linear filter. Under the assumption that the signal estimate is contaminated
by foreground residuals induced by parameterizable systematic effects,
we infer the systematics-induced contamination by cross-correlating
the initial signal and foreground estimates. Correcting for the inferred error,
we are able to 
subtract foreground contamination from the linearly filtered 
signal up to the first order in the amplitude of the telescope systematics. In simulations of an interferometric 21-cm survey,
our algorithm removes foreground leakage induced by complex gain errors by one to two orders
of magnitude in the power spectrum.
Our technique thus eases the requirements on 
telescope characterization for modern and next-generation 21-cm cosmology 
experiments.
\end{abstract}

\maketitle

\section{introduction}
\label{sec:introduction}

The 21-cm line in neutral hydrogen (HI) has emerged as a new and highly-promising
tool in
cosmology. By observing the cumulative hydrogen signal 
from many unresolved sources, \emph{hydrogen intensity mapping}
\citep{chang_pen, Loeb} uses HI as a tracer of matter to survey 
large volumes of the universe rapidly. 
In particular, this technique can
map large-scale structure in the intermediate/low redshift 
universe ($z < 4$) \citep{chang_pen}, constrain ionization fraction 
and reionization models during the epoch of reionization 
($ z \sim 6 - 10$) \citep{FURLANETTO2006181}, and potentially observe 
matter distribution throughout much of the dark ages ($z > 30$)
\citep{Zaldarriaga_2004}. The promise of this technique is reflected 
by the numerous 21-cm cosmology experiments which are either collecting 
data (GMRT~\citep{Choudhuri_2020}, 
HERA~\citep{HERA_2017a}, LOFAR~\citep{LOFAR_2017}, 
MWA~\citep{MWA_2019}, CHIME~\citep{chime_overview_2021}) or 
being planned (HIRAX~\citep{2016SPIE.9906E..5XN},
CHORD~\citep{2019clrp.2020...28V}, PUMA~\citep{2019BAAS...51g..53S}, 
SKA~\citep{SKA}).

However, detecting the 21-cm signal is difficult due to astrophysical
foregrounds which are $\sim$3--5 orders of magnitude brighter, primarily consisting
of Galactic synchrotron emission
and extra-galactic point sources \citep{ssp+14}. Using the smooth spectral shape of 
these foregrounds, many filters have been proposed to separate the 21-cm 
signal from foregrounds. These filters can be categorized either as
linear or non-linear methods. 
Linear filters, such as those based on the Karhunen-Loève 
(KL) eigenmode projection \citep{COBE_KL} or the delay filter \citep{delay_filter},
can remove foregrounds so long as the instrument response is 
accurately known or well behaved, 
but they are highly sensitive to telescope systematic errors 
(\textsl{eg.} calibration errors), leaving foreground residuals that dominate 
the signal \citep{all_sky_richard}. On the other hand, 
non-linear filtering methods such as principal
component analysis (PCA) have historically been more successful 
in single-dish telescope experiments to detect the 21-cm signal in cross-correlation
with galaxy surveys 
\citep{2010Natur.466..463C,2013ApJ...763L..20M,2018MNRAS.476.3382A, 2022MNRAS.510.3495W}. 
However, non-linear filters are difficult to characterize
in general and may result in over-subtraction and signal loss.
The recent detection of the 21-cm large-scale structure signal by CHIME used a
foreground filter that is linear apart from a final flagging stage, however
an aggressive delay filter
that eliminated the largest spacial scales was required
\citep{2022arXiv220201242C}.

In this paper, we introduce a hybrid foreground filtering technique,
where the data are initially processed by a linear filter to obtain initial
signal and foreground estimates.
We then cross-correlate the foreground and signal channels in
order to draw out and isolate systematics-induced residual foreground
contamination in the latter. This second step is non-linear, however, 
in contrast to other non-linear methods, allows
for a perturbative expansion (in powers of the magnitude of the systematic errors and
the signal-to-foreground ratio) to control the non-linearity. 
Thus, signal loss through the procedure can be characterized analytically.

Critical to this procedure is the choice of \emph{how} to cross-correlate the signal
and foreground estimates, \textsl{i.e.} through what transformations---or in which
data subspaces---should this cross-correlation be performed. We address this question by
framing the correlation as a quadratic estimator for (small) parameters describing the
systematic errors. An example of such parameters are errors in the complex gains of
the signal chains of an interferometric array. This interferometric calibration problem
has been a central focus of the 21-cm literature, and is the case on which we focus in
this work. In simulations of a small, compact, square 
antenna array, we show that our hybrid 
foreground filtering algorithm can suppress calibration-error induced foreground 
residuals in the power-spectrum by two orders of magnitude compared to the
linear filter alone. The algorithm thus dramatically eases the requirements on telescope
calibration for 21-cm surveys. We also discuss the extension of this technique to 
other types of systematics commonly 
observed in 21-cm experiments.

We first illustrate the basic idea of the hybrid foreground filtering 
algorithm using a toy example in Section~\ref{sec:toy_example}.
We provide a general 
formalism of the technique in Section~\ref{sec:formalism}. 
In Section~\ref{sec:simulation_pipeline}, we give a 
summary of the simulation pipeline that we use to test our algorithm. 
In Section~\ref{sec:examples}, we apply the hybrid 
technique to simulated data in three 
different scenarios, each with systematics of increasing complexity, 
and show the results. Finally, in Section~\ref{sec:discussion}, 
we discuss the current 
limitations of our hybrid algorithm and its
extension to 
other types of systematic errors commonly 
observed in 21-cm experiments. We also compare this hybrid technique
with other gain calibration 
and foreground removal techniques in literature.
We present the conclusions in Section~\ref{sec:conclusions}.

\section{Toy example}
\label{sec:toy_example}

We introduce the technique with a simple yet illustrative example.
Suppose we have a data set
$\vect{d}$ which represents a sky map with $N$ 
frequencies and $M$ pixels per 
frequency, with $N, M \gg 1$. The sky map includes both the 21-cm signal 
and foregrounds, with $\vect{s}$ and $\vect{f}$ representing the 
vectorized version of each component. We expect the foregrounds
to be spectrally smooth. In this example, we will assume that $\vect{f}$
is independent of frequency. The model for the data is then
\begin{equation}
\label{eq:d}
    d_{\nu p} = s_{\nu p} + f_{p},
\end{equation}
where $f_{p} = \left(\vect{f}\right)_{(\nu p)}$, reflecting the assumption that foregrounds are frequency-independent.  The indices 
$\nu$ and $p$ represent the frequency channel and pixel number, respectively, and form a single compound index for 
the vectors $\vect{d}$, $\vect{s}$, and $\vect{f}$.
Suppose that
$\langle\vect{f}\rangle, \langle\vect{s}\rangle=\vect{0}$,
and that signal and foregrounds are uncorrelated with covariances
\begin{equation}
\label{eq:S_F_def}
\begin{split}
    & S_{(\nu p)(\nu^\prime p^\prime)} = 
    \langle s_{(\nu p)} s_{(\nu^\prime p^\prime)} \rangle =
    \delta_{\nu \nu^\prime} \delta_{p p^\prime} \sigma_s^2,  \\
    & F_{(\nu p)(\nu^\prime p^\prime)} = 
    \langle f_{p} f_{p^\prime} \rangle =
    \delta_{p p^\prime} \sigma_f^2,  \\ 
    & \text{with } \sigma_f^2 \gg \sigma_s^2.
\end{split}
\end{equation}

Equations.~\eqref{eq:d} and \eqref{eq:S_F_def} represent our simple model for foregrounds
that are completely correlated in frequency and the cosmological
signal that is uncorrelated in frequency and position.
We want to separate $\vect{s}$ and $\vect{f}$ given the data $\vect{d}$. Since
the foregrounds are independent of frequency, we can estimate
$\vect{f}$ by averaging the data over frequency. Namely, we have
\begin{equation} \label{intro_f_hat}
    \hat{f}_p = \frac{1}{N} \sum_\nu d_{\nu p} ,
\end{equation}
where $\hat{f}_p$ represents the estimated foreground $\vect{\hat{f}}$ at pixel $p$.
(Note that the dimensions of $\vect{\hat{f}}$ and $\vect{f}$ 
are different since $\vect{\hat{f}}$ is obtained after averaging over the frequency axis.)
We can represent the
operation of frequency averaging by a matrix $\matr{A}$ whose elements are
\begin{equation}
\label{eq:A}
   A_{(p)(\nu^\prime p^\prime)} = \frac{1}{N}
   \delta_{p p^\prime}.
\end{equation}

Eq. \eqref{intro_f_hat} can now be written as
\begin{equation}
\label{eq:f_hat}
\begin{split}
   \vect{\hat{f}} = \matr{A}\vect{d} 
   \hspace{0.25in} \longrightarrow \hspace{0.25in}
   &\hat{f}_p = f_p + \frac{1}{N} \sum_{\nu^\prime} s_{\nu^\prime p},   \\
   & \langle \hat{f}_p \hat{f}_{p^\prime} \rangle = 
    \left( \sigma_f^2 + \frac{\sigma_s^2}{N} \right) \delta_{p p^\prime}.
\end{split}
\end{equation}

With foregrounds estimated, we can obtain the estimated signal $\vect{\hat{s}}$ simply by subtracting the estimated foregrounds from the data. This operation can be represented by a matrix $\matr{K}=\matr{I}-\matr{A}^\prime$, where $\matr{A}^\prime$ is the same filter as $\matr{A}$, except that it maintains the original dimensions of the data. Namely,  
\begin{equation}
\label{eq:A_prime}
   A^\prime_{(\nu p)(\nu^\prime p^\prime)} = \frac{1}{N}
   \delta_{p p^\prime}.
\end{equation}
Then, the matrix $\matr{K}$ is 
\begin{equation}
\label{eq:K_toy}
   K_{(\nu p)(\nu^\prime p^\prime)} = 
   \delta_{p p^\prime}\left(\delta_{\nu \nu^\prime} - \frac{1}{N} \right).
\end{equation}

The signal estimate $\vect{\hat{s}}$ is obtained by
\begin{equation}
\label{eq:s_hat}
\begin{split}
   \vect{\hat{s}} = \matr{K}\vect{d} 
   \hspace{0.25in} \longrightarrow \hspace{0.25in}
   &\hat{s}_{\nu p} =  s_{\nu p} - \frac{1}{N} \sum_{\nu^\prime} s_{\nu^\prime p}.
\end{split}
\end{equation}
Note that the signal estimate $\vect{\hat{s}}$ is a linear combination of the components of
the true signal $\vect{s}$ only, so the foreground filter $\matr{K}$ effectively 
separates signal from foregrounds.

Let us now examine what happens to signal and foreground estimates when we 
introduce band-pass (independent of pixel) perturbations to the data. 
In this case, the data model is
\begin{equation}
\label{eq:d_pert}
    d_{\nu p} = \left(s_{\nu p} + f_{p}\right)\left( 1 + g_\nu \right),
    \hspace{0.3in} g_\nu \ll 1.
\end{equation}

Using $\matr{A}$ and $\matr{K}$ from Eqs.~\eqref{eq:f_hat} and \eqref{eq:s_hat}, the signal and foreground estimates now
become
\begin{eqnarray}
\label{eq:f_s_hat_pert}
\begin{split}
   \hat{f}_p =& f_p \left( 1 + \frac{1}{N}\sum_{\nu^\prime} g_{\nu^\prime} \right) + \frac{1}{N} \sum_{\nu^\prime} 
   s_{\nu^\prime p} \left( 1 + g_{\nu^\prime} \right), \\
   \hat{s}_{\nu p} =&  s_{\nu p} \left( 1 + g_{\nu} \right)
       - \frac{1}{N} \sum_{\nu^\prime} 
       s_{\nu^\prime p}\left( 1 + g_{\nu^\prime} \right) + \\
        &f_p \left( g_\nu - \frac{1}{N}\sum_{\nu^\prime} g_{\nu^\prime} \right).
\end{split}
\end{eqnarray}
Eqs.~\eqref{eq:f_s_hat_pert} show how band pass perturbations
cause foregrounds leaking into the 
initially signal dominated subspace
(the last term in the equation for $\hat{s}_{\nu p}$).
The power in the estimated signal is 
now\footnote{Unless stated otherwise, $\langle \cdot \rangle $ denotes
ensemble averaging over signal and foreground realizations
while keeping the perturbations fixed.}
\begin{equation}
\label{eq:s_hat_var}
\begin{split}
   \langle \hat{s}_{\nu p}^2 \rangle = &
       \sigma_s^2 \left[ \left(1-\frac{2}{N} \right)\left( 1 + g_{\nu} \right)^2 + \frac{1}{N^2}\sum_{\nu^\prime} \left( 1 + g_{\nu^\prime} \right)^2 \right] + \\
       & \sigma_f^2 \left( g_\nu - \frac{1}{N}\sum_{\nu^\prime} g_{\nu^\prime} \right)^2.
\end{split}
\end{equation}
Equation~(\ref{eq:s_hat_var}) shows that perturbations introduce a relative bias in the
signal power estimate by a term of order 
$(\sigma_g\sigma_f/\sigma_s)^2$, where $\sigma_g$
represents the scale of the gain perturbations. 
This factor is typically much greater than one and as such, our signal
estimate is dominated by systematic-error induced residual foregrounds.
This is the essence of the problem we aim to address in this paper.

From Eq.~\eqref{eq:f_s_hat_pert} and the assumption that foregrounds are much brighter than the signal, we expect that $\hat{f}_p$ goes
roughly as $\sim f_p$ and that $\hat{s}_{\nu p}$ goes roughly as
$\sim g_\nu f_p$.
Thus, one way to estimate the perturbations $g_\nu$ is by cross correlating 
the signal and foreground estimates. 
In particular, let us define the estimates $\hat{y}_\nu$  
\begin{equation}
\label{eq:y_hat}
\begin{split}
    \hat{y}_\nu = \frac{\sum_{p} \hat{f}_p \hat{s}_{\nu p}}
                  {\sum_{p} \hat{f}_p^2}.
\end{split}             
\end{equation}

If we plug the expressions for $\hat{f}_p$ and $\hat{s}_{\nu p}$ from Eq.~\eqref{eq:f_s_hat_pert}
into Eq. \eqref{eq:y_hat}, take the average, and use
Eq. \eqref{eq:S_F_def} to simplify, we find
\begin{equation}
\label{eq:y_hat_mean}
\begin{split}
    \langle \hat{y}_\nu \rangle =& 
        \left(\displaystyle g_\nu - \frac{1}{N}\sum_{\nu^\prime} g_{\nu^\prime}\right) \left (
        1 + \frac{1}{N}\sum_{\nu^\prime} g_{\nu^\prime}\right)^{-1}\\
        &+ \mathcal{O}\left(\frac{\sigma_s^2 g}{N\sigma_f^2}\right)\\
        = &\sum_{\nu^\prime} W_{\nu \nu^\prime} g_{\nu^\prime} 
        + \mathcal{O}\left(\frac{\sigma_g^2}{N}\right)
\end{split}             
\end{equation}
\noindent where the window matrix $\matr{W}$ is defined as
\begin{equation}
\label{eq:W}
\begin{split}
    W_{\nu \nu^\prime} = 
        \delta_{\nu \nu^\prime} - \frac{1}{N}.
\end{split}             
\end{equation}

Rather than inverting $\matr{W}$ to obtain a first-order estimate for
$g_\nu$,
we note that, for this particular example, the combination
$\langle \hat{y}_\nu \rangle \hat{f}_p$ gives the foreground
term that we need to remove from $\hat{s}_{\nu p}$ in 
Eq.~\eqref{eq:f_s_hat_pert} 
(this will not be the case in general).
Thus, we define the `cleaned' signal as
\begin{equation}
\label{eq:s_tilde}
    \tilde{s}_{\nu p} = \hat{s}_{\nu p} - \hat{y}_\nu \hat{f}_p.
\end{equation}

With the expressions for $\hat{f}_p$, $\hat{s}_{\nu p}$, and 
$\hat{y}_\nu$ given in Eqs.~\eqref{eq:f_s_hat_pert}, and 
\eqref{eq:y_hat}, it is shown in Appendix~\ref{app:toy_model}
that the cleaned signal 
$\tilde{s}_{\nu p}$ has zero mean
and variance
\begin{equation}
\label{eq:s_tilde_var}
\begin{split}
   \langle \tilde{s}_{\nu p}^2 \rangle =
       \sigma_s^2 \left[1 + \mathcal{O}\left(\sigma_g, \frac{1}{M}\right) \right].
\end{split}
\end{equation}
where $\mathcal{O}\left(\sigma_g, 1/M\right)$
means that the next terms in the expansion are of order $\sigma_g$ and $1/M$.
On average, the cleaned signal is free of foreground bias to all orders.
Thus, in the case of the toy example, the foreground residual subtraction is exact. In general, the foreground residual subtraction may not be complete and can leave behind higher order foreground terms in the cleaned power spectrum.

\section{Formalism}
\label{sec:formalism}

In Section \ref{sec:toy_example}, we have shown that if linear filters $\matr{A}$ and $\matr{K}$ 
can produce foreground and signal estimates, then the cross-correlation in Eq.~\eqref{eq:y_hat} can estimate the perturbations. In fact, this procedure can be generalized and framed as a quadratic estimator. 

To see this explicitly, we can define a matrix $\matr{E}_\nu$ for each frequency $\nu$ as
\begin{equation}
\label{eq:E_nu}
    \left({E}_\nu\right)_{(p^\prime)(\nu^{\prime\prime} p^{\prime\prime})} = \delta_{p^\prime p^{\prime\prime}} \delta_{\nu \nu^{\prime\prime}},
\end{equation}
Equation~\eqref{eq:y_hat} can now be written as
\begin{equation}
\label{eq:y_hat1}
\begin{split}
    \hat{y}_\nu &= \frac{1}{\eta}\vect{\hat{f}}^\dagger \matr{E}_\nu \vect{\hat{s}} - b_\nu,
\end{split}             
\end{equation}
where $b_\nu = 0$, and 
\begin{equation} \label{eq:toy_normalization}
    \eta = \sum_{p} \hat{f}_p^2 = \vect{\hat{f}}^\dagger\vect{\hat{f}}
\end{equation}
is a normalization factor.
Equation~\eqref{eq:y_hat1} is reminiscent of the optimal quadratic
estimator formalism developed in \cite{tegmark97b}.
In this formalism, $\matr{E}_\nu$ can be an arbitrary symmetric matrix
and $b_\nu$ is chosen accordingly to make the
estimator unbiased. There are two important differences, however. The first is the appearance of the normalization factor $\eta$ which is computed from filtered data. The second difference is that in our case we are not correlating
data with itself. Instead, we are correlating two different
vectors, or more precisely, two linear 
transformations of the original data set. Another way to see this difference
with the traditional quadratic estimator is that if we re-write
$\hat{y}_\nu$ as
\begin{equation}
\label{eq:y_hat2}
\begin{split}
    \hat{y}_\nu &=  \frac{1}{\eta}\vect{d}^\dagger \matr{A}^\dagger \matr{E}_\nu \matr{K} \vect{d} - b_\nu,
\end{split}             
\end{equation}
then we see that $\matr{A}^\dagger \matr{E}_\nu \matr{K}$
is not necessarily symmetric.

To develop the toy example into a more general formalism (where we allow frequency-dependent foregrounds and any type of parametrizable systematic errors), we write the perturbed data as
\begin{equation} \label{eq:d_pert1}
    \vect{d} = (\matr{I} + \matr{G}) (\vect{s} + \vect{f}),
\end{equation}
where $\matr{I}$ is the identity matrix, and $\matr{G}$ is a perturbation matrix that assigns errors to the data. Given a set of perturbations $\{g_i\}$, we can parameterize $\matr{G}$ as
\begin{equation} \label{eq:formalism_pert_matrix}
    \matr{G} = \sum_i g_i \matr{\Gamma}_i,
\end{equation}
where $\matr{\Gamma}_i$ are base matrices that represent how different perturbations act on data.
For instance, in the case of the toy example, we can add band-pass error to the data by defining $\matr{G}$ as
\begin{equation} \label{eq:G_nu}
    \mbf{G} = \sum_{\nu} g_\nu \mbf{\Gamma}_\nu,
\end{equation}
where the matrix $\matr{\Gamma}_\nu$ is defined as
\begin{equation} \label{formalism_gain_matrix_2}
    (\Gamma_\nu)_{(\nu^\prime p^\prime) (\nu^{\prime\prime} p^{\prime\prime})} = I_{(\nu^\prime p^\prime)(\nu^{\dprime}p^{\dprime})}\delta_{\nu \nu^\prime} = \delta_{\nu^\prime \nu^{\prime \prime}} \delta_{p^\prime p^{\prime \prime}} \delta_{\nu \nu^\prime}.
\end{equation}
In this case, the base matrix $\matr{\Gamma}_\nu$ corresponding to the $\nu$-th frequency is the identity matrix 
with diagonal elements corresponding to other frequencies set to zero. In other words, $\matr{\Gamma}_\nu$ picks out all the data points that are corrupted by the band-pass gain $g_\nu$.

Note that the data model in Eq.~(\ref{eq:d_pert1}) should
also include a noise term $\vect{n}$ reflecting the fact that our data will
also contain instrumental noise. For simplicity, we omit the noise
term in this section since the statistics of $\vect{n}$ are expected to be similar
to those of $\vect{s}$, and thus the foreground filter will act on both components in 
the same way. However, noise is included in the simulations presented
in Sections~\ref{sec:simulation_pipeline} and \ref{sec:examples}.

Applying a linear foreground filter $\matr{K}$ to the data, we obtain the estimated signal
\begin{equation} \label{eq:form_s_hat}
\begin{split}
    \vect{\hat{s}} = \matr{K} \vect{d} & = \matr{K}(\matr{I} + \matr{G})(\vect{s} + \vect{f}) \\
    & = \matr{K}(\matr{I} + \matr{G})\vect{s} + \matr{K}\matr{G}\vect{f} \approx \matr{K}\matr{G}\vect{f},
\end{split}
\end{equation}
assuming the foreground filter works well in absence of systematics, i.e. $\matr{K}\vect{f} \ll \vect{s}$. Note that Eq.~(\ref{eq:form_s_hat}) is dominated by the foreground residual term $\matr{K}\matr{G}\vect{f}$. On the other hand, when we apply the signal filter $\matr{A}$ to data, we get the estimated foreground
\begin{equation} \label{eq:form_f_hat}
\begin{split}
    \vect{\hat{f}} = \matr{A} \vect{d} & = \matr{A}(\matr{I} + \matr{G})(\vect{s} + \vect{f}) \\
    & \approx \vect{f} ,
\end{split}
\end{equation}
assuming $\vect{f} \gg \vect{s}$ and the elements of the perturbation matrix $\matr{G}$ are small.

Since $\vect{\hat{s}}$ is dominated by terms at the order of $\matr{G}\vect{f}$, and $\vect{\hat{f}}$ goes roughly as $\vect{f}$, we can cross correlate the two vectors to estimate the perturbations, namely elements of the matrix $\matr{G}$. For each perturbation $g_i$, we can define its estimate as
\begin{equation} \label{eq:pert_estimate}
    \hat{y}_i = \frac{\vect{\hat{f}}^\dagger \matr{E}_i \hat{\vect{s}}}{\vect{\hat{f}}^\dagger \matr{D}_i \vect{\hat{f}}} - b_i,
\end{equation}
where $\matr{E}_i$ is a quadratic estimator, and $\matr{D}_i$ is the normalization operator which controls how normalization is done [note that $\matr{D}_i$ is the identity matrix in Eq. (\ref{eq:toy_normalization})]. We will comment on the choice of $\matr{E}_i$ and  $\matr{D}_i$ in the last two paragraphs of this section.

To first order in the amplitude of the perturbations, we can express the ensemble average of the perturbation estimate $\hat{y}_i$ as a linear combination of the true perturbations:
\begin{equation} \label{eq:y_hat_exp}
    \langle \hat{y}_i \rangle = \sum_{i^\prime} W_{i i^\prime} g_{i^\prime} - b_i.
\end{equation}

To determine the window matrix $\matr{W}$, it is useful to first compute the variance of the data $\vect{d}$ and the covariance between the estimated signal $\vect{\hat{s}}$ and estimated foreground $\vect{\hat{f}}$. With $\vect{d}$ defined in Eq. (\ref{eq:d_pert1}), we can compute
\begin{equation} \label{eq:data_cov}
\begin{split}
    \mbf{C}^{dd} & = \langle \bm{d} \bm{d}^\dagger \rangle \\
    & \approx \mbf{S} + \mbf{F} + \mbf{G}(\mbf{S} + \mbf{F}) + (\mbf{S} + \mbf{F})\mbf{G}^\dagger \\
    & = \matr{S} + \matr{F} + \sum_{i} \left[ g_{i} \matr{\Gamma}_i
    \left( \matr{S} + \matr{F}\right)  +
    g_{i}^* 
    \left( \matr{S} + \matr{F}\right) \matr{\Gamma}_i^\dagger
    \right].
\end{split}
\end{equation}
Note that we have dropped second order terms in perturbations on the second line of Eq. (\ref{eq:data_cov}). The covariance between $\vect{\hat{s}}$ and $\vect{\hat{f}}$ is now
\begin{equation} \label{eq:sf_cov}
    \matr{C}^{sf} \equiv \langle \vect{\hat{s}} \vect{\hat{f}}^{\dagger} \rangle = \matr{K} \matr{C}^{dd} \matr{A}^\dagger \approx \sum_{i} g_{i} \matr{\Phi}_i,
\end{equation}
where
\begin{equation} \label{eq:phi}
    \matr{\Phi}_i = \frac{\partial 
    \matr{C}^{\vect{\hat{s}}\vect{\hat{f}}}}{\partial g_i} \approx
    \matr{K} \matr{\Gamma}_i \matr{F}\matr{A}^\dagger.
\end{equation}

To compute the window matrix, we take the ensemble average of Eq.~(\ref{eq:pert_estimate}) while keeping perturbations fixed and assume $b_i = 0$ (we can check this assumption by examining whether the perturbation estimate is biased at the end of the calculation). We obtain
\begin{equation} \label{start_to_compute_W}
    \langle \hat{y}_i \rangle = \left\langle \frac{\vect{\hat{f}}^\dagger \matr{E}_i \hat{\vect{s}}}{\vect{\hat{f}}^\dagger \matr{D}_i \vect{\hat{f}}} \right\rangle.
\end{equation}
We can approximate the right-hand side of Eq.~(\ref{start_to_compute_W}) by the ratio between the expectation of the numerator and that of the denominator. With details shown in Appendix~\ref{sec:formalism_app}, we find that Eq.~(\ref{start_to_compute_W}) to the first order in amplitude of the perturbations gives
\begin{equation} \label{eq:window_result}
    \langle\hat{y}_i \rangle = \sum_{i^\prime} \frac{\text{Tr}(\matr{E}_i\matr{\Phi}_{i^\prime})}{\text{Tr}[\matr{A}^\dagger \matr{D}_i \matr{A} (\matr{S} + \matr{F})]}g_{i^\prime} + \frac{\text{Tr}[\matr{E}_i\matr{K}(\matr{S} + \matr{F})\matr{A}^\dagger]}{\text{Tr}[\matr{A}^\dagger \matr{D}_i \matr{A} (\matr{S} + \matr{F})]}.
\end{equation}

Comparing Eq.~(\ref{eq:window_result}) with Eq.~(\ref{eq:y_hat_exp}), we see that
\begin{equation}
\begin{split}
    W_{i i^\prime} & = \frac{\text{Tr}(\matr{E}_i\matr{\Phi}_{i^\prime})}{\text{Tr}[\matr{A}^\dagger \matr{D}_i \matr{A} (\matr{S} + \matr{F})]} \\
    & \approx \frac{\text{Tr}(\matr{E}_i\matr{K} \mbf{\Gamma}_{i^\prime}\mbf{F}\matr{A}^\dagger)}{\text{Tr}[\matr{A}^\dagger \matr{D}_i \matr{A} \matr{F}]} \label{eq:formalism_W},
\end{split}
\end{equation}
where we have simplified the expression on the second line using Eq.~(\ref{eq:phi}) and $\matr{AS} \ll \matr{AF}$. In addition, to make the perturbation estimate unbiased, we need to set $b_i$ in Eq.~(\ref{eq:y_hat_exp}) to be
\begin{equation}
    b_i = \frac{\text{Tr}[\matr{E}_i\matr{K}(\matr{S} + \matr{F})\matr{A}^\dagger]}{\text{Tr}[\matr{A}^\dagger \matr{D}_i \matr{A} (\matr{S} + \matr{F})]}.
\end{equation}
However, given that $\matr{K}\matr{F}$ and $\matr{S}\matr{A}^\dagger$ are approximately 0, $b_i$ can be neglected (and $b_i$ is exactly 0 if we choose $\matr{K}$ and $\matr{A}$ to be orthogonal projections). For this reason, we will set $b_i = 0$ in the rest of this paper.

To recover perturbations $g_i$ from $\hat{y}_i$, we need to compensate for the window $\matr{W}$. In principle, this can be done by inverting the window matrix, but $\matr{W}$ is ill-conditioned because the linear filters $\matr{K}$ and $\matr{A}$ have removed some modes from the data. We can nonetheless partially recover perturbations from the remaining modes with the pseudo-inverse of the window matrix, $\mbf{W}^{+}$. Denoting the recovered perturbations by $\hat{g}_i$, we have
\begin{equation} \label{eq:recovered_gain}
    \hat{g}_i = \sum_{i^\prime} W^+_{i i^\prime} \hat{y}_{i^\prime}.
\end{equation}
Although we cannot recover the perturbations perfectly, we expect that the missing modes in perturbations are not needed since those are modes already removed by the foreground filter. We can then assemble the recovered perturbation matrix $\mbf{\hat{G}}$ in the same way that $\matr{G}$ is constructed:
\begin{equation} \label{eq:formalism_G_hat}
    \mbf{\hat{G}} = \sum_{i} \hat{g}_i \mbf{\Gamma}_i.
\end{equation}

Equation~(\ref{eq:form_s_hat}) shows the foreground contamination term in the estimated signal is $\mbf{K}\mbf{G}\bm{f}$. We can now reconstruct this term using the estimates and subtract it from the estimated signal to obtain the cleaned signal. Since $\mbf{\hat{G}}$ approximates true perturbations up to the first order, when we apply the filter $\matr{K}$ to the recovered perturbations, we have:
\begin{equation}
    \mbf{K\hat{G}} = \mbf{KG} + \mathcal{O}(\bm{G}^2).
\end{equation}
We also already have a foreground estimate $\vect{\hat{f}}$. Equation~(\ref{eq:form_f_hat}) shows that $\vect{\hat{f}} \approx \vect{f} + \mathcal{O}(\matr{G}\vect{f})$. Therefore, we can reconstruct the contamination term up to the first order in amplitude of perturbations using $\matr{K}$, $\mbf{\hat{G}}$, and $\vect{\hat{f}}$:
\begin{equation}
    \mbf{K}\mbf{\hat{G}}\bm{\hat{f}} = \mbf{K}\mbf{G}\bm{f} + \mathcal{O}(\mbf{G}^2\bm{f}).
\end{equation}
Now we subtract this term from $\vect{\hat{s}}$ to obtain the cleaned signal, denoted by $\vect{\tilde{s}}$:
\begin{equation} \label{eq:cleaned_s}
    \bm{\tilde{s}} = \bm{\hat{s}} - \mbf{K}\mbf{\hat{G}}\bm{\hat{f}} = \mbf{K}(\matr{I}+\matr{G})\bm{s} + \mathcal{O}(\mbf{G}^2\bm{f}).
\end{equation}
In Eq.~(\ref{eq:cleaned_s}), we see that as long as perturbations are small enough so that $\mbf{G}^2\bm{f} < \bm{s}$, the cleaned signal $\bm{\tilde{s}}$ will no longer be overwhelmed by foreground residuals.

Note that Eqs.~(\ref{eq:d_pert1}),~(\ref{eq:formalism_pert_matrix}), and~(\ref{eq:form_s_hat})-(\ref{eq:cleaned_s}) are general
and can be applied to any linear filter $\matr{A}$ and $\matr{K}$, foreground and signal covariance model $\matr{F}$ and $\matr{S}$, perturbation model $\matr{G}$, and choice of quadratic estimator $\matr{E}_i$ and normalization operator $\matr{D}_i$. It can be verified that if we substitute the quantities in the formalism with those defined in our toy example [Eqs.~(\ref{eq:S_F_def}), ~(\ref{eq:A}),~(\ref{eq:K_toy}),~(\ref{eq:E_nu}),~(\ref{eq:toy_normalization}), and~(\ref{eq:G_nu})], then we obtain the results
for $\matr{W}$ and $b_\nu$ in Eqs. \eqref{eq:W} and \eqref{eq:y_hat1}.

In principle, both the quadratic estimator $\matr{E}_i$ and normalization operator $\matr{D}_i$ can be chosen arbitrarily, but not all choices will result in good estimates for $\hat{y}_i$. In Eq. \eqref{eq:E_nu}, the quadratic estimator $\matr{E}_\nu$ is chosen to recover Eq.~\eqref{eq:y_hat} in the toy example. We could instead follow the optimal quadratic estimator formalism developed by Tegmark \citep{tegmark97b} to define $\matr{E}_\nu$ such that it minimizes the variance $\langle (\hat{y}_v)^2 \rangle$ and determine the normalization operator $\matr{D}_\nu$ accordingly. 


However, the optimal quadratic estimator formalism requires assumed signal and foreground models, whereas we would prefer our estimator to be model independent. In addition, for illustrative purposes, we prefer the simplicity of estimators similar to Eq.~(\ref{eq:E_nu}) in order to demonstrate the hybrid foreground filtering technique. For these reasons, we leave the optimal quadratic estimator formalism for future studies and will stay with intuitive choices of $\matr{E}_i$ in this paper. 

One particular choice for the quadratic estimator and normalization operator is to set them equal to the perturbation base matrix
\begin{equation} \label{eq:E_i_choise}
    \matr{E}_i = \matr{D}_i = \matr{\Gamma}_i.
\end{equation}
Recall that the base matrix $\matr{\Gamma}_{i}$ picks up all the data that are corrupted by the $i$-th perturbation. It is an intuitive choice because in order to estimate the $i$-th perturbation $g_i$, we naturally want to cross-correlate all the data points that are affected by $g_i$ while leaving out the rest. This is essentially what we have done in the toy model. Note that in Eq.~(\ref{eq:y_hat}), we only cross-correlate the estimated signal at the $\nu$-th frequency with the estimated foreground when estimating the bandpass error $g_\nu$. The motivation for this choice of the quadratic estimator may seem naive, but we will demonstrate its applicability to estimating complex gain errors in Section~\ref{sec:examples} using simulations outlined in the following section.



\section{Summary of the simulations}
\label{sec:simulation_pipeline}

We now provide an overview of the simulations used to test our hybrid algorithm. A more in-depth description is provided in Appendix \ref{app:pipeline}. In this paper, we use a KL-based foreground filter to fulfill the role of $\matr{K}$ in Eq. (\ref{eq:form_s_hat}). The KL method requires prior knowledge of the sky components encoded in the covariance matrices of the signal $\matr{S}$ and foregrounds $\matr{F}$. For simplicity, we compute the prior covariance matrices from simulated Monte Carlo (MC) realizations given a simple angular power spectrum and frequency dependency of each component.  However, we adopt an independent and  more realistic sky model based on  \cite{hdb+18} and \cite{odb+18} to generate input maps as our test data set in order to verify the robustness of the foreground removal algorithms to foreground model. Previous work has found that as long as the models of the sky components that are input to the KL filter are qualitatively correct, the filter is insensitive to the exact model mismatch between it and the test data \citep{Shaw_2014}.

We consider four components in the sky model: (1) cosmological HI signal, (2) synchrotron radiation from cosmic ray electrons gyrating in Galactic magnetic fields, (3) free-free radiation due to free electrons scattering off ions, and (4) extra-galactic point sources. Both the simple prior simulations and more realistic test data sets of these components are described in Appendix \ref{sec:HIsim} to \ref{sec:p_source}. 
The simulated telescope is a $5\times5$ square array consisting of 6~m aperture single dishes with 1~m separation in between any adjacent pair of dish edges. All the antennas have the same primary beam derived from a fixed antenna illumination pattern. The telescope has a system temperature of 50 K, observing from 400 MHz to 500 MHz with 50 evenly spaced frequency channels. The total integration time of the observation is 120 days. (In Section \ref{subsec:bandpass_error} and \ref{subsec:antenna_error_unstacked}, the survey observes the same patch of the sky for 120 days, but in Section \ref{subsec:antenna_error}, the survey observes 15 different sky patches for 8 days each.) The simulated telescope takes in the more realistic sky maps and generates visibilities through a Fourier transform under the flat sky approximation, which are then corrupted by systematic effects and noise. Details on the instrument and visibility generation are included Appendix \ref{instr}. After visibilities are computed, we apply the hybrid foreground filter developed in Section \ref{sec:formalism} to remove foreground residuals and recover the HI signal.

To quantify the performance of our foreground filter, we compare the power spectra of the recovered HI signal with the theoretical power spectrum used to generate the input HI maps. The power spectrum estimator is constructed using the optimal quadratic estimator formalism \citep[e.g.,][]{dod03} and computes the redshift-averaged spectrum of the given recovered HI signal. We will refer readers to Appendix \ref{sec:ps_estimator} for details on the power spectrum estimator.

As a first step to develop and test the hybrid foreground filtering algorithm, we have made several assumptions and simplifications in the simulations. We only consider complex gain errors and omit others, such as beam or baseline errors, in the simulated telescope (although we will address how to extend the foreground filtering algorithm to handle these errors in Section \ref{subsec:other_systematics}). In Section \ref{subsec:bandpass_error} and \ref{subsec:antenna_error_unstacked}, we assume that we can integrate on one patch of sky for 120 days without the baselines rotating, while in Section \ref{subsec:antenna_error}, we simulate a time axis by simply adding uncorrelated patches of the sky. These simplifications facilitate implementation of simulations while preserving key aspects of real surveys required to demonstrate the algorithm. 

\section{Examples of applying hybrid foreground filtering}
\label{sec:examples}

In this section, we will demonstrate the hybrid foreground filter through three examples, starting with the simplest case of band-pass perturbations, and finishing with a more complex case of antenna and frequency dependent gain perturbations.

\subsection{Band-pass perturbations}
\label{subsec:bandpass_error}

The data product of the simulated telescope is visibilities. Since band-pass errors are independent of baselines, we only consider visibilities from non-redundant baselines---which we will refer to as the stacked visibilities---by averaging visibilities of redundant baselines. Visibilities are represented by the vector $(\bm{v_d})_{(\nu b)}$, with $\nu$ and $b$ indexing frequencies and non-redundant baselines, respectively. Following the same format as Eq.~(\ref{eq:d_pert1}), the data is a sum of HI and foreground visibilities, denoted by $\bm{v_{HI}} \text{ and } \bm{v_F}$ respectively, and is multiplied by the perturbation matrix $\mbf{G}$:
\begin{equation} \label{v_d}
    \bm{v_d} = (\mbf{I} + \mbf{G}) (\bm{v_{HI}} + \bm{v_F}).
\end{equation}
Since the real-valued band-pass error is already considered in the toy model, we can define $\mbf{G}$ by modifying Eq.~(\ref{eq:G_nu}) and~(\ref{formalism_gain_matrix_2}):
\begin{eqnarray}
\begin{aligned}
    \mbf{G} &= \sum_{\nu} g_\nu \mbf{\Gamma}_\nu,  \label{gain_matrix} \\
    (\Gamma_\nu)_{(\nu^\prime b^\prime) (\nu^{\prime\prime} b^{\prime\prime})} &= \delta_{\nu^\prime \nu^{\prime \prime}} \delta_{b^\prime b^{\prime \prime}} \delta_{\nu \nu^\prime}, \label{gamma_matrix_ex1}
\end{aligned}
\end{eqnarray}
where $g_\nu$'s are the band-pass errors that we want to estimate. Note that we have replaced the index $p$ in Eq.~(\ref{formalism_gain_matrix_2}) with $b$ in Eq.~(\ref{gamma_matrix_ex1}) to be consistent with using visibilities as data instead of the sky map. 

Having defined the data format, we now apply the hybrid foreground filtering algorithm following the procedure developed in Section~\ref{sec:formalism}. We first obtain the estimated signal $\bm{\hat{v}_{HI}}$ by applying the KL filter $\mbf{K}$ to the data:
\begin{equation} \label{v_HI_hat}
\begin{split}
    \bm{\hat{v}_{HI}} = \mbf{K}\bm{v_d}. 
\end{split}
\end{equation}
As in the toy example, we choose the filter $\mbf{A}$ to be $\mbf{I} - \mbf{K}$ and obtain the estimated foreground:
\begin{equation} \label{v_F_hat}
    \bm{\hat{v}_F} = (\mbf{I} - \mbf{K}) \bm{v_d}. 
\end{equation}

We now estimate the gain error by cross-correlating the estimated foreground with the estimated signal. Adopting the choice we made for the quadratic estimator and normalization operator in Eq.~(\ref{eq:E_i_choise}), we get
\begin{eqnarray}
    \matr{E}_\nu = \matr{D}_\nu = \matr{\Gamma}_\nu, \label{eq:E_band_pass}
\end{eqnarray}
Substitute $\bm{\hat{v}_{HI}}$, $\bm{\hat{v}_F}$, $\matr{E}_\nu$, and $\matr{D}_\nu$ into Eq.~(\ref{eq:pert_estimate}), we obtain the perturbation estimate

\begin{eqnarray}
\begin{aligned}
     \hat{y}_\nu 
      & = \frac{\bm{\hat{v}_{F}}^\dagger\mbf{\Gamma}_\nu \bm{\hat{v}_{HI}}}{\bm{\hat{v}_F}^\dagger\mbf{\Gamma}_\nu \bm{\hat{v}_F}} \label{ex1y_hat} \\
      & = \frac{\sum_{b} (\hat{v}_{F})^{*}_{(\nu b)} (\hat{v}_{HI})_{(\nu b)}}{\sum_{b} (\hat{v}_{F})^{*}_{(\nu b)} (\hat{v}_{F})_{(\nu b)}} \label{ex1y_hat_ind}.
\end{aligned}
\end{eqnarray}

We can see the similarity between Eq.~(\ref{ex1y_hat}) and Eq.~(\ref{eq:y_hat}) of the toy model.
Both estimate perturbations by averaging the product of the estimated foreground and estimated signal over the index which perturbations are independent of.

Recall that the ensemble average of the perturbation estimate $\hat{y}_\nu$ is a linear combination of the actual perturbations $g_{\nu^\prime}$
\begin{equation} \label{eq:ex1_linear_comb}
    \langle \hat{y}_\nu \rangle = \sum_{\nu^\prime} W_{\nu \nu^\prime} g_{\nu^\prime}.
\end{equation}
Using Eq.~(\ref{eq:formalism_W}),~(\ref{eq:E_band_pass}), and $\matr{A} = \matr{I} - \matr{K}$, we obtain the window matrix
\begin{equation} \label{window}
\begin{split}
    W_{\nu \nu^\prime}  = \frac{\text{Tr} [\mbf{\Gamma}_\nu \mbf{K} \mbf{\Gamma}_{\nu^\prime} \mbf{F} (\mbf{I} - \mbf{K}^\dagger)]}{\text{Tr} \left[  (\mbf{I} - \mbf{K}^\dagger) \mbf{\Gamma}_\nu (\mbf{I} - \mbf{K})\mbf{F}  \right]}.
\end{split}
\end{equation}

Note that the derivation of the window matrix in Section~\ref{sec:formalism} does not account for noise in the data. However, since the KL filter mostly preserves the noise as it does to the signal, noise terms will propagate just like the signal throughout the derivation. The effect of noise is therefore negligible assuming the noise covariance is much smaller than the foreground covariance, given a reasonable amount of integration time. 

To test the accuracy of the window matrix, we draw 50 band-pass errors from a Gaussian distribution with a standard deviation of $10^{-3}$ and apply them to the simulated telescope. Following the steps summarized in section IV, we simulate telescope observations of a single sky patch for 120 days. We then obtain the estimated errors $\hat{y}_\nu$ using equation (\ref{ex1y_hat}) from the visibilities and compare them with the true errors $g_\nu$ as well as true errors passed through the window matrix, namely $\sum_{\nu^\prime} W_{\nu \nu^\prime} g_{\nu^\prime}$, which approximates $\langle \hat{y}_\nu \rangle$ in Eq.~(\ref{eq:ex1_linear_comb}).

The result is shown in Fig.~1. Note that although the estimated errors $\hat{y}_\nu$ roughly trace the actual errors $g_\nu$, they do not match exactly since the estimated errors are in fact a linear combination of the true errors as seen in Eq.~(\ref{eq:ex1_linear_comb}). This also explains why the estimated errors better match the true errors passed through the window matrix. The small deviation between the two comes from the fact that the gain $\hat{y}_\nu$ is only estimated from one data set rather than an ensemble of realizations.

\begin{figure*}
\includegraphics[scale=0.28]{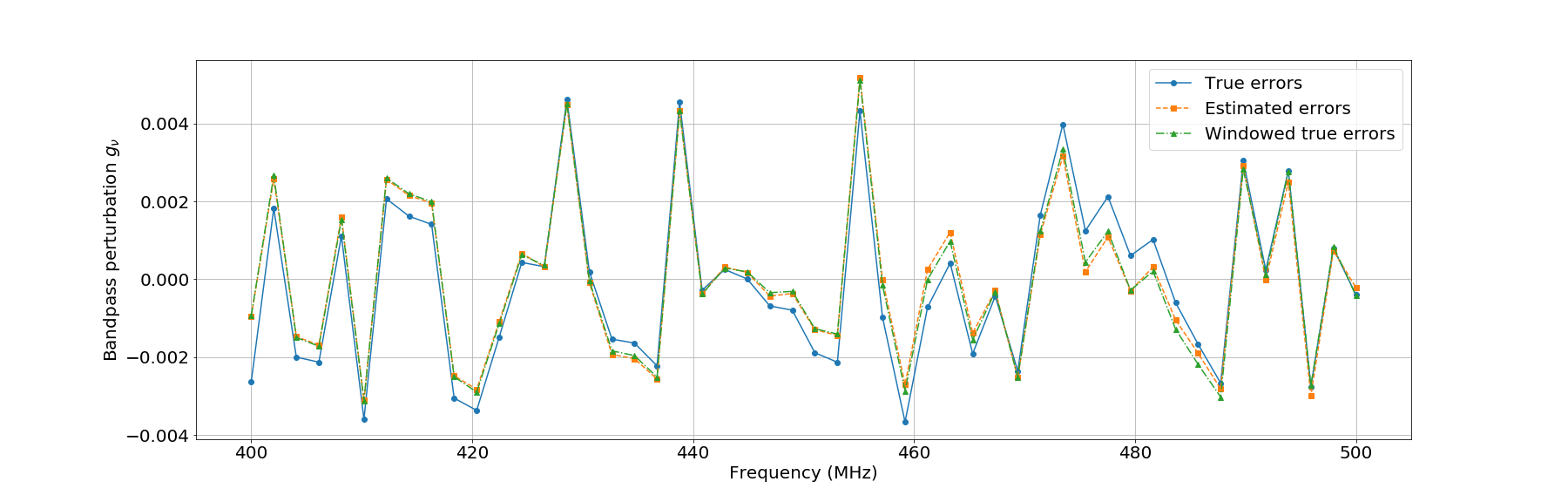}
\caption{\label{gain} Simulated band-pass errors and recovery using our quadratic estimator. The band-pass errors are drawn from a Gaussian distribution with a standard deviation of $10^{-3}$ and are subsequently estimated using our simple quadratic estimator. The estimated errors match the true errors passed through the window matrix. This verifies the accuracy of the window matrix and shows our method is able to recover band-pass errors with good precision.}
\end{figure*}

To obtain the recovered errors $\hat{g}_\nu$, we need to compensate for the window matrix. As discussed in Section~\ref{sec:formalism}, we cannot invert the window matrix since it is ill-conditioned, but we can partially recover the errors using the pseudo-inverse:
\begin{equation} \label{g_hat}
    \hat{g}_\nu = \sum_{\nu^\prime} W^+_{\nu \nu^\prime} \hat{y}_{\nu^\prime}.
\end{equation}
We then assemble the recovered perturbation matrix $\mbf{\hat{G}}$ as in Eq.~(\ref{gain_matrix}):
\begin{equation}
    \mbf{\hat{G}} = \sum_{\nu} \hat{g}_\nu \mbf{\Gamma}_\nu.
\end{equation}
Following Eq.~(\ref{eq:cleaned_s}), we can now obtain the cleaned signal $\bm{\tilde{v}_{HI}}$ by subtracting the foreground contamination term from the estimated signal
\begin{equation} \label{v_HI_tilde}
    \bm{\tilde{v}_{HI}} = \bm{\hat{v}_{HI}} - \mbf{K}\mbf{\hat{G}}\bm{\hat{{v}}_F}.
\end{equation}

\begin{figure*}
\includegraphics[scale=0.28]{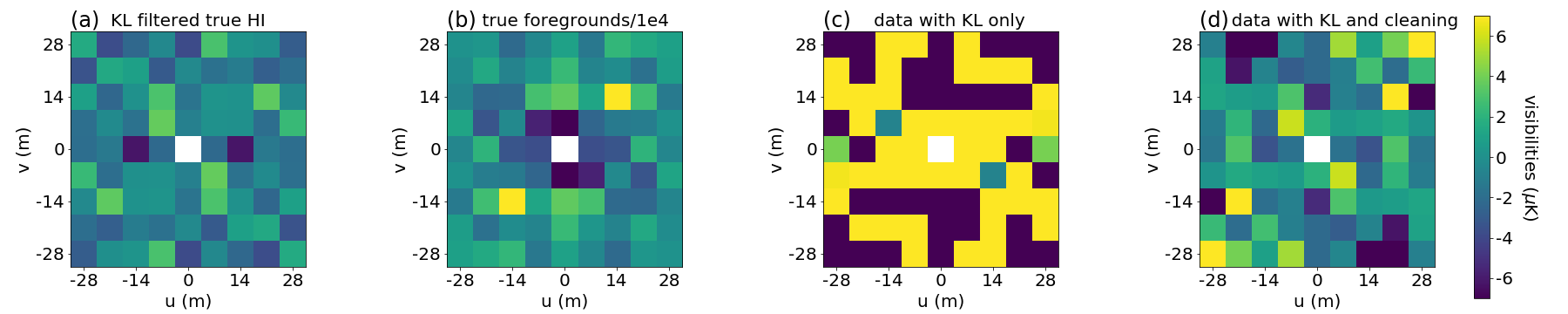}
\caption{\label{visibilities} Visibilities of signal, foregrounds, contaminated signal, and recovered signal in the $uv$ plane at 400 MHz: (a) true HI visibilities passed through the KL filter, (b) true foreground visibilities passed through the KL filter and suppressed by $10^{4}$, (c) visibilities of the sky map (foregrounds + HI + noise) with $10^{-3}$ level band-pass errors after KL filtering but before foreground residual subtraction, and (d) previous visibilities but after foreground subtraction. Only the real part of the visibilities is shown. In comparison to panel (c), panel (d) contains much fewer over-saturated visibilities. This suggests the recovered signal is no longer dominated by foreground contamination after foreground residual subtraction.}
\end{figure*}

We expect the cleaned signal $\bm{\tilde{v}_{HI}}$ to be no longer dominated by foreground residuals if $\mbf{G}^2\bm{v_{F}} < \bm{v_{HI}}$ as indicated by Eq.~(\ref{eq:cleaned_s}). Our simulation shows that foregrounds are brighter than the HI signal by roughly 5 orders of magnitude [see panel (a) and (b) of Fig.~\ref{visibilities}]. This suggests foreground subtraction can remove foreground contamination due to band-pass errors up to the order of $10^{-3}$. Panel (c) and (d) of Fig.~\ref{visibilities} compare sky visibilities in the $uv$ plane before and after foreground residual subtraction. Both visibilities contain $10^{-3}$-level band-pass errors and have been passed through the KL filter. Note that before foreground residual subtraction, visibilities are mostly over-saturated due to foreground contamination. After the subtraction, most visibilities are comparable with HI signal [panel (a)] in terms of magnitude.

\begin{figure}[h!]
\includegraphics[scale=0.34]{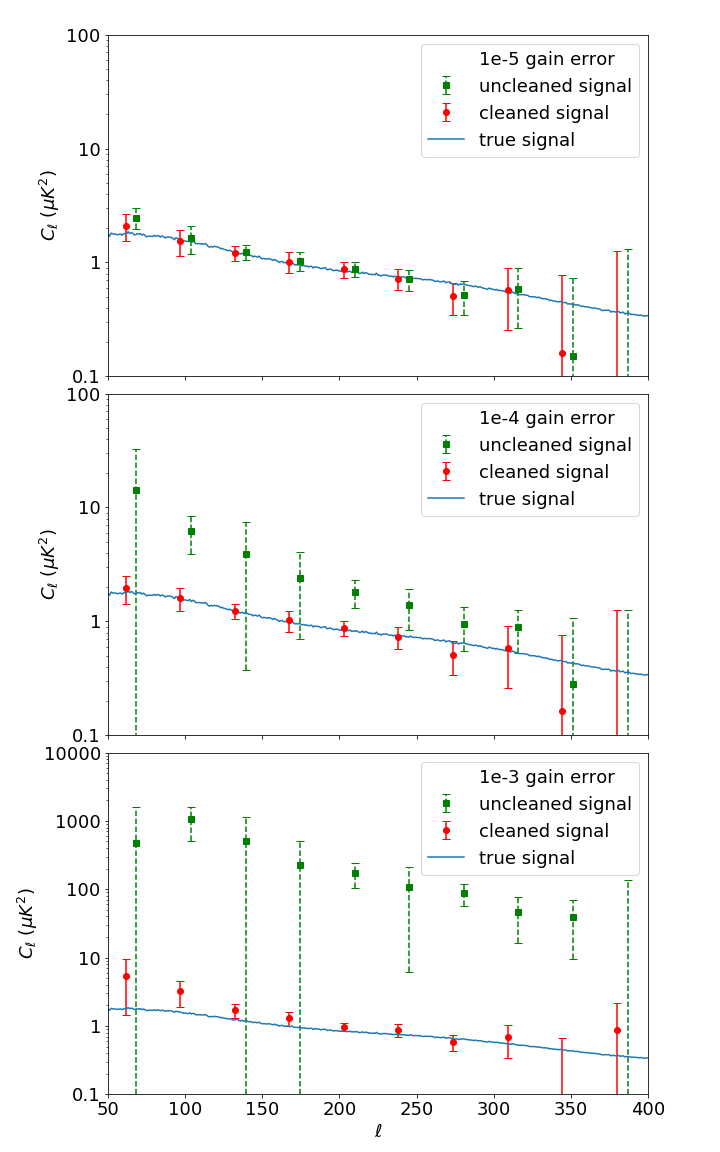}
\caption{\label{first_case_fig} Comparison of the HI power of the uncleaned signal (with KL filtering only), cleaned signal (with KL filtering and foreground residual subtraction), and true signal. The top, middle, and bottom panels correspond to band-pass errors at the order of $10^{-5}$, $10^{-4}$, and $10^{-3}$, respectively. At the $10^{-5}$ level, foreground contamination is negligible. At the $10^{-4}$ level, the HI power of the uncleaned data shows a bias which foreground residual subtraction is able to remove. At the $10^{-3}$ level, the HI power of the uncleaned data shows a larger bias, but that of the cleaned data only shows slight bias at low $l$. This suggests the foreground residual subtraction can suppress foreground contamination due to band-pass errors by nearly three orders of magnitude.}
\end{figure}

We now use the quadratic estimator outlined in Section~\ref{sec:simulation_pipeline} to estimate HI power spectrum from the visibilities. We compare the power spectrum of the linearly-filtered signal $\vect{\hat{v}_{HI}}$ and that of the signal cleaned with our algorithm $\vect{\tilde{v}_{HI}}$ in the presence of band-pass errors drawn from a Gaussian distribution, depicted in the top, middle, and bottom panels of Fig.~\ref{first_case_fig}, with standard deviations of $10^{-5}$, $10^{-4}$, and $10^{-3}$, respectively. The blue curve in the three panels is the theoretical HI power spectrum used to generate the data. 

Figure~\ref{first_case_fig} shows that at the $10^{-5}$ level, band-pass errors are too small to cause a bias in the HI power even before foreground residual subtraction. At the $10^{-4}$ level, however, the HI power of the uncleaned signal is one order of magnitude higher than the theoretical power, but foreground residual subtraction is able to remove this bias in the HI power of the cleaned signal. At the $10^{-3}$ level, the HI power of the uncleaned shows a bias of three orders of magnitude while that of the cleaned signal only shows slight bias at low $l$. This indicates the foreground residual subtraction can effectively suppress foreground contamination due to band-pass errors at the order of $10^{-3}$ or below.

\subsection{Antenna-dependent perturbations with a time axis}
\label{subsec:antenna_error}

Now we want to generalize band-pass errors from the previous example to antenna-dependent complex gain errors. In the simulated telescope, we model the gain of the $i$-th antenna at the $\nu$-th frequency as
\begin{equation} \label{ex2_err_model}
    1 + q_{\nu i} = (1 + h_\nu + p_i + \delta_{\nu i})e^{2\pi i (\nu \tau_i + \epsilon_{\nu i})},
\end{equation}
where $q_{\nu i}$ is the gain error which has contributions from a band-pass error $h_\nu$, antenna dependent gain error $p_i$, random gain error $\delta_{\nu i}$, delay error $\tau_i$, and random phase error $\epsilon_{\nu i}$. Note that the delay error has the unit of microsecond when the frequency is given in MHz, while other error components are unitless. 

In this example, we will estimate antenna-dependent errors from stacked visibilities. Since the stacked visibility averages visibilities of redundant baselines, the total gain error $g_{(\nu b)}$ of the stacked visibility at frequency $\nu$ and baseline $b$ is the averaged sum of errors from every antenna pair $i$ and $j$ that forms the baseline $b$:
\begin{equation}
    g_{(\nu b)} = \frac{1}{N(b)} \sum_{(i, j) \in b} q_{\nu i} + q^*_{\nu j},
\end{equation}
where $N(b)$ is the total number of antenna pairs with baseline b, and we have assumed $q_{\nu i}, q_{\nu j} \ll 1$. Then the perturbation matrix, defined in equation (\ref{eq:formalism_pert_matrix}), now becomes
\begin{equation} \label{ex2_gain_matrix}
    \mbf{G} = \sum_{\nu, b} g_{(\nu b)} \mbf{\Gamma}_{\nu b}, 
\end{equation}
and
\begin{equation}
\begin{split}
(\Gamma_{\nu b})_{(\nu^\prime b^\prime)(\nu^{\prime\prime} b^{\prime\prime})}
& = (\Gamma_\nu)_{(\nu^\prime b^\prime)(\nu^{\prime\prime} b^{\prime\prime})} \delta_{b b^\prime},
\end{split}
\end{equation}
where $(\Gamma_\nu)_{(\nu^\prime b^\prime)(\nu^{\prime\prime} b^{\prime\prime})}$ is the perturbation base matrix defined in Eq.~(\ref{gamma_matrix_ex1}) from the previous example. The extra term $\delta_{b b^\prime}$ reflects the fact that perturbations in the current example also depend on baselines, so the matrix $\matr{\Gamma}_{\nu b}$ not only picks up visibilities with the frequency $\nu$ but also baseline $b$ at the same time.

Compared with the previous band-pass case, antenna-dependent errors have many more parameters to estimate, so more information needs to be included in the data to compensate for the larger parameter space. We can achieve this by adding a time axis to the stacked visibilities, and we will denote all quantities in this expanded space by capital caligraphic letters:
\begin{equation} \label{eq:visibility_ex2}
\begin{split}
    & (v_{HI})_{(\nu b)} \rightarrow (\mathcal{V}_{HI})_{(\nu b t)}, \\
    & (v_{F})_{(\nu b)} \rightarrow (\mathcal{V}_{F})_{(\nu b t)}.
\end{split}
\end{equation}
The time axis represents the telescope observing different patches of the sky at different times. In our simulations, the telescope observes 15 sky patches with an integration time of 8 days each [so the subscript $t$ in Eq.~(\ref{eq:visibility_ex2}) ranges from 1 to 15 in this case]. Since different sky patches do not correlate, the signal and foreground covariances in the expanded space can be related to the original signal and foreground covariances by
\begin{equation} \label{eq:ex2_S_F}
\begin{split}
    & \mathcal{S}_{(\nu b t) (\nu^\prime b^\prime t^\prime)} = S_{(\nu b) (\nu^\prime b^\prime)} \delta_{t t^\prime}, \\
    & \mathcal{F}_{(\nu b t) (\nu^\prime b^\prime t^\prime)} = F_{(\nu b) (\nu^\prime b^\prime)} \delta_{t t^\prime}.
\end{split}
\end{equation}
Similarly, the KL filter estimates the signal of one sky patch by only using information from the same patch, so the KL filter in the expanded space is
\begin{equation} \label{eq:ex2_K}
    \mathcal{K}_{(\nu b t) (\nu^\prime b^\prime t^\prime)} = K_{(\nu b) (\nu^\prime b^\prime)} \delta_{t t^\prime}.
\end{equation}

We make the assumption that antenna-dependent gain errors $g_{(\nu b)}$ are constant with respect to time. Then, the perturbation matrix $\bm{\mathcal{G}}$ is related to the original perturbation matrix defined in equation (\ref{ex2_gain_matrix}) by:
\begin{equation}
\begin{split}
    \mathcal{G}_{(\nu^\prime b^\prime t^\prime) (\nu^{\prime\prime} b^{\prime\prime} t^{\prime\prime})} & = G_{(\nu^\prime b^\prime) (\nu^{\prime\prime} b^{\prime\prime})} \delta_{t^\prime t^{\prime\prime}} \\
    & = \sum_{\nu, b} g_{(\nu b)} (\Gamma_{\nu b})_{(\nu^\prime b^\prime) (\nu^{\prime\prime} b^{\prime\prime})} \delta_{t^\prime t^{\prime\prime}} \\
    & = \sum_{\nu, b} g_{(\nu b)} (\Delta_{\nu b})_{(\nu^\prime b^\prime t^\prime) (\nu^{\prime\prime} b^{\prime\prime} t^{\prime\prime})},
\end{split}
\end{equation}
where we have defined the base matrix in the expanded space as
\begin{equation} \label{eq:ex2_delta_matrix}
    (\Delta_{\nu b})_{(\nu^\prime b^\prime t^\prime) (\nu^{\prime\prime} b^{\prime\prime} t^{\prime\prime})} = (\Gamma_{\nu b})_{(\nu^\prime b^\prime) (\nu^{\prime\prime} b^{\prime\prime})} \delta_{t^\prime t^{\prime\prime}}.
\end{equation}

Now we can define the data, estimated signal, and estimated foreground in the same way as in Eqs.~(\ref{v_d}),~(\ref{v_HI_hat}), and~(\ref{v_F_hat}), respectively, with relevant qualities changed to their counterparts in the expanded space:
\begin{equation} \label{eq:ex2_data_estimate}
\begin{split}
    & \bm{\mathcal{V}_d} = (\bm{\mathcal{I}} + \bm{\mathcal{G}}) (\bm{\mathcal{V}_{HI}} + \bm{\mathcal{V}_F}), \\
    & \bm{\hat{\mathcal{V}}_{HI}} = \bm{\mathcal{K}}\bm{\mathcal{V}_d}, \\
    & \bm{\hat{\mathcal{V}}_F} = \bm{\mathcal{V}_d} - \bm{\hat{\mathcal{V}}_{HI}}.
\end{split}
\end{equation}

Using the newly defined base matrix $\matr{\Delta}_{\nu b}$ in Eq.~(\ref{eq:ex2_delta_matrix}), the quadratic estimator and normalization operator chosen in Eq.~(\ref{eq:E_i_choise}) now become
\begin{eqnarray}
    \bm{\mathcal{E}}_{\nu b} = \bm{\mathcal{D}}_{\nu b} = \matr{\Delta}_{\nu b}, \label{eq:E_ex2}
\end{eqnarray}
Substituting $\bm{\mathcal{E}}_{\nu b}$, $\bm{\mathcal{D}}_{\nu b}$, and other quantities in the expanded space into Eq.~(\ref{eq:pert_estimate}), we obtain the perturbation estimate
\begin{equation} \label{ex2_y_hat}
\begin{split}
     \hat{y}_{(\nu b)} & = \frac{\bm{\hat{\mathcal{V}}_{F}}^\dagger(\bm{\Delta}_{\nu b}) \bm{\hat{\mathcal{V}}_{HI}}}{\bm{\hat{\mathcal{V}}_F}^\dagger(\bm{\Delta}_{\nu b}) \bm{\hat{\mathcal{V}}_F}} \\
     & = \frac{\sum_{t} (\bm{\hat{\mathcal{V}}_{F}})^{*}_{(\nu b t)} (\bm{\hat{\mathcal{V}}_{HI}})_{(\nu b t)}}{\sum_{t} (\bm{\hat{\mathcal{V}}_{F}})^{*}_{(\nu b t)} (\bm{\hat{\mathcal{V}}_{F}})_{(\nu b t)}}.
\end{split}
\end{equation}
Compare Eq.~(\ref{ex2_y_hat}) with the perturbation estimate of the band-pass case from Eq.~(\ref{ex1y_hat}), we see that instead of summing over the baseline, we now sum over the time axis, i.e., always summing over the axis over which the gain errors are constant. This is, in fact, a feature of the quadratic estimator designed in Eq.~(\ref{eq:E_i_choise}).

Using Eq.~(\ref{eq:formalism_W}), we obtain the window matrix
\begin{equation}
    W_{(\nu b)(\nu^\prime b^\prime)} = \frac{\text{Tr} [\bm{\Delta}_{\nu b} \bm{\mathcal{K}} \bm{\Delta}_{\nu^\prime b^\prime} \bm{\mathcal{F}} (\matr{I} - \bm{\mathcal{K}}^\dagger)]}{\text{Tr} \left[ (\bm{\mathcal{I}} - \bm{\mathcal{K}}^\dagger) \bm{\Delta}_{\nu b} (\bm{\mathcal{I}} - \bm{\mathcal{K}}) \bm{\mathcal{F}}  \right]}.
\end{equation}
Following Eq.~(\ref{eq:recovered_gain}),~(\ref{eq:formalism_G_hat}), and~(\ref{eq:cleaned_s}), we recover the error by compensating the window
\begin{equation}
    \hat{g}_{(\nu b)} = \sum_{\nu^\prime, b^\prime} W^+_{(\nu b) (\nu^\prime b^\prime)} \hat{y}_{(\nu^\prime b^\prime)}
\end{equation}
and recover the perturbation matrix
\begin{equation}
    \bm{\hat{\mathcal{G}}} = \sum_{\nu, b} \hat{g}_{(\nu b)} \bm{\mathcal{E}}_{(\nu b)}.
\end{equation}
We finally obtain the cleaned signal
\begin{equation}
    \bm{\tilde{\mathcal{V}}_{HI}} = \bm{\hat{\mathcal{V}}_{HI}} - \bm{\mathcal{K}}\bm{\hat{\mathcal{G}}}\bm{\hat{\mathcal{V}}_F}.
\end{equation}

We use the simulations developed in Section~\ref{sec:simulation_pipeline} again to test foreground residual subtraction in the case of antenna-dependent gain errors. The sky maps now include a time axis which has 15 realizations. This simulates the telescope observing 15 different sky patches at different times. We keep the total integration time 120 days, so each of the 15 sky patches is observed for 8 days. We start by comparing the HI power spectrum of the KL filtered data before and after foreground residual subtraction with error components $h_\nu$, $p_i$, $\delta_{\nu i}$, $\tau_i$, and $\epsilon_{\nu i}$ of Eq.~(\ref{ex2_err_model}) each at the $10^{-5}$ level (accounting for the factor of $2\pi$ multiplied with $\epsilon_{\nu i}$ and frequency multiplied with $\tau_i$). We then repeat the analysis by redrawing all error components with their standard deviations increased by 10 times and then again with their standard deviations increased by 100 times. 

\begin{figure}[h!]
\includegraphics[scale=0.34]{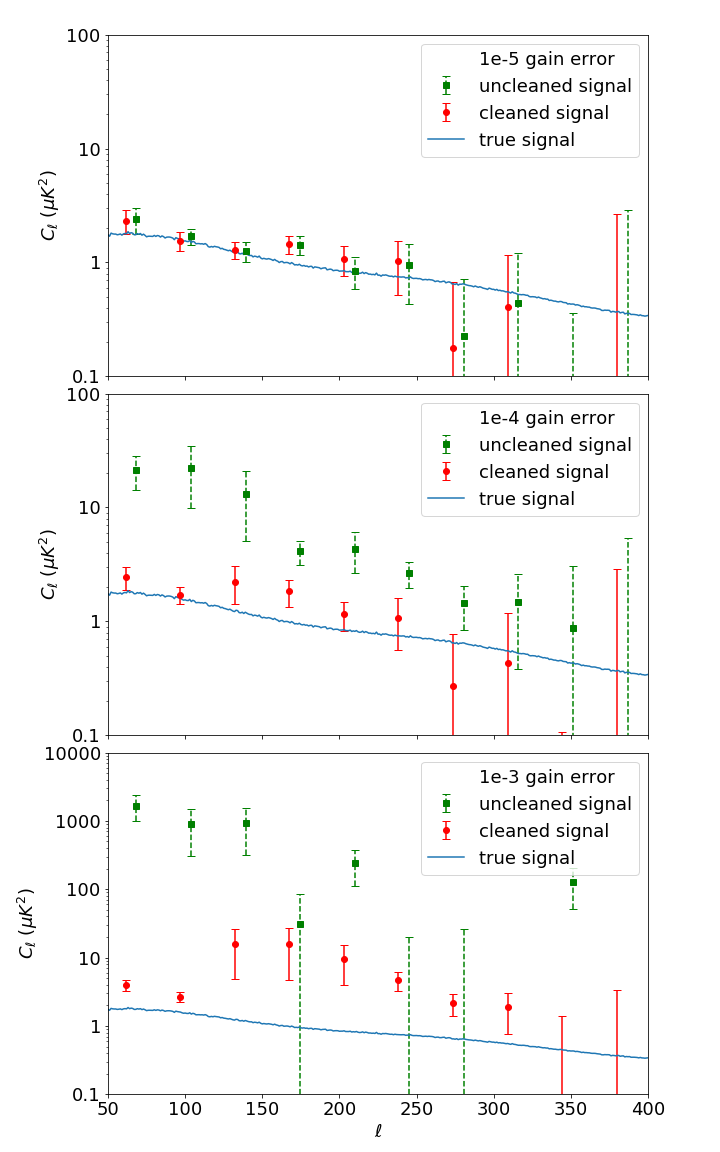}
\caption{\label{second_case_fig} HI power spectrum of the uncleaned signal (with KL filtering only), cleaned signal (with KL filtering and foreground residual subtraction), and true signal in the presence of antenna-dependent errors and a time axis. The top, middle, and bottom panels correspond to each component of the error at the order of $10^{-5}$, $10^{-4}$, and $10^{-3}$, respectively. At the $10^{-5}$ level, errors are too small to cause foreground bias even before the foreground residual subtraction. At the $10^{-4}$ level, HI power of the uncleaned signal shows a one-order-of-magnitude bias while HI power of the cleaned signal shows only slight bias. At the $10^{-3}$ level, foreground residual subtraction is able to suppress foreground bias in the HI power by two orders of magnitude.}
\end{figure}

The results are shown in Fig.~\ref{second_case_fig}. The top, middle, and bottom panel correspond to individual error components of Eq.~(\ref{ex2_err_model}) at the order of $10^{-5}$, $10^{-4}$, and $10^{-3}$, respectively. Similar to the band-pass case, $10^{-5}$-level errors do not cause significant foreground contamination. At the $10^{-4}$ level, the HI power of the uncleaned signal shows a bias of one order of magnitude, but HI power after foreground residual subtraction matches the theoretical power (shown in blue). At the $10^{-3}$ level, the power spectrum of the uncleaned signal shows roughly a three-order-of-magnitude bias while the HI power after foreground residual subtraction is only biased by one order of magnitude. 

Compared with Fig.~\ref{first_case_fig}, HI power of the cleaned signal in Fig.~\ref{second_case_fig} has larger bias in the middle and bottom panels. This is not surprising because antenna-depend errors have multiple components, so the rms errors in the antenna-depend case are larger than those in the band-pass case. Nonetheless, Fig.~\ref{second_case_fig} suggests that the foreground residual subtraction is able to remove foreground bias for antenna-depend errors up to the order of $10^{-4}$.

\subsection{Antenna-dependent perturbations estimated in the unstacked space}
\label{subsec:antenna_error_unstacked}

In the previous section we assumed that antenna-dependent gains are roughly constant for all 15 observations, which is not always valid. To address this limitation, we now apply foreground residual subtraction to the observation of a single sky patch. We will do so in the unstacked visibility space i.e., considering all the redundant baselines. Even though redundant visibilities contain identical information of the sky, they carry additional information about which antenna pairs produce the visibilities. 

We denote the visibility produced by the $i$th and $j$th antenna at the $\nu$-th frequency as $v_{(\nu i j)}$. It is a sum of HI signal and foregrounds multiplied with antenna errors
\begin{equation}
\begin{split}
    &v_{(\nu i j)} \\
    & = \left(1 + q_{\nu i}\right) \left(1 + q^*_{\nu j}\right) \left[(v_{HI})_{(\nu i j)} + (v_{F})_{(\nu i j)}\right] \\
    & \approx \left(1 + q_{\nu i} + q^*_{\nu j}\right) \left[(v_{HI})_{(\nu i j)} + (v_{F})_{(\nu i j)}\right],
\end{split}
\end{equation}
where the antenna-dependent error $q_{\nu i}$ was defined in Eq.~(\ref{ex2_err_model}), and we assume $q_{\nu i},~q_{\nu j} \ll 1$. Also, notice that the second antenna in the pair (antenna $j$ in this case) has its errors complex conjugated.

Given this data, we can design the intuitive quadratic estimator in the same way as Eq.~(\ref{ex1y_hat_ind}) and~(\ref{ex2_y_hat}) to estimate the perturbation $q_{\nu i}$. Namely, we cross-correlate the foreground estimate $(\hat{v}_{F})_{(\nu i j)}$ with the signal estimate $(\hat{v}_{HI})_{(\nu i j)}$ by summing their product over the index $j$, of which the error $q_{\nu i}$ is independent. We thus have
\begin{equation} \label{ex3_y_hat}
    \hat{y}_{\nu i} = \frac{\sum_{j} (\bm{\hat{v}_F})^*_{(\nu i j)}(\bm{\hat{v}_{HI}})_{(\nu i j)} }{\sum_{j} (\bm{\hat{v}_{F}})^*_{(\nu i j)} (\bm{\hat{v}_F})_{(\nu i j)}}.
\end{equation}

Recall that the estimated signal is dominated by foreground residuals (i.e., foregrounds multiplied with gain error) and that the estimated foregrounds are dominated by the true foregrounds. Then, Eq.~(\ref{ex3_y_hat}) becomes 
\begin{eqnarray}
    \hat{y}_{\nu i} \sim \frac{\sum_j \left(q_{\nu i} + q^*_{\nu j}\right) \left|(\bm{v_F})_{(\nu i j)}\right|^2}{\sum_j \left|(\bm{v_F})_{(\nu i j)}\right|^2} \\ = q_{\nu i} + \frac{\sum_j q^*_{\nu j} \left|(\bm{v_F})_{(\nu i j)}\right|^2}{\sum_j \left|(\bm{v_F})_{(\nu i j)}\right|^2}. \label{eq:ex3_intuitive_estimator}
\end{eqnarray}
Therefore, the perturbation estimate $\hat{y}_{\nu i}$ will be mainly sensitive to $q_{\nu i}$ if the second term of Eq.~(\ref{eq:ex3_intuitive_estimator}) is small. This is only when the errors of the $j$th antenna are uncorrelated, but the antenna errors in our model have a correlated component: the band-pass error. As a result, the second term of Eq.~(\ref{eq:ex3_intuitive_estimator}) obtains a large contribution from the band-pass error, requiring us to compensate for the window matrix to demix the errors. 

Equation~(\ref{eq:ex3_intuitive_estimator}) suggests that we need to disentangle both the antenna errors and their complex conjugates to compensate for the window. One way to do this is to regard the complex conjugates of the gain perturbations as independent parameters. We define $q_{\nu (N_a+i)} \equiv q^*_{\nu i}$ for all $i$ from 1 to $N_a$, where $N_a$ is the total number of antennae. Namely, we treat the complex conjugate of the $i$th error as the $(N_a+i)$-th error. 
With this numerical trick we treat the problem as if there were doubled the number of parameters, however, the number of degrees of freedom and information content has not changed since the visibilities $v_{(\nu i j)}$ and $v_{(\nu j i)}$ contain identical information.

With this change, the visibilities are now effectively produced by $2N_a$ antennae but with the $1$st to $N_a$-th antenna only appearing first in an antenna pair and the $(N_a+1)$-th and $(2N_a)$-th antenna only appearing second in a pair, i.e., the first antenna index running from $1$ to $N_a$ and the second antenna index running from $N_a+1$ to $2N_a$. We will refer to this visibility space as the redundancy unstacked space. We can modify Eq.~(\ref{ex3_y_hat}) accordingly in order to estimate the $i$th antenna error:
\begin{equation} \label{ex3_y_hat_1}
    \hat{y}_{\nu i} = \frac{\displaystyle\sum_{j=N_a+1, j\neq N_a+i}^{2N_a} (\bm{\hat{v}_F})^*_{(\nu i j)} (\bm{\hat{v}_{HI}})_{(\nu i j)}}{\displaystyle\sum_{j=N_a+1, j\neq N_a+i}^{2N_a} (\bm{\hat{v}_{F}})^*_{(\nu i j)} (\bm{\hat{v}_F})_{(\nu i j)}} \mbox{    } \text{if $i = 1$ to $N_a$}
\end{equation}
and 
\begin{equation} \label{ex3_y_hat_2}
\begin{split}
    &\hat{y}_{\nu i} = \frac{\displaystyle\sum_{j=1, j \neq i - N_a}^{N_a} (\bm{\hat{v}_F})^*_{(\nu j i)}(\bm{\hat{v}_{HI}})_{(\nu j i)}}{\displaystyle\sum_{j=1, j \neq i - N_a}^{N_a} (\bm{\hat{v}_{F}})^*_{(\nu j i)} (\bm{\hat{v}_F})_{(\nu j i)}} \mbox{    } \\
    &\text{if $i = N_a+1$ to $2N_a$,}
\end{split}
\end{equation}
where the restrictions $j \neq N_a+i$ in Eq.~(\ref{ex3_y_hat_1}) and $j \neq i - N_a$ in Eq.~(\ref{ex3_y_hat_2}) come from the fact that we do not include auto-correlations in the data.

Having determined the data format and chosen the quadratic estimator, we can now apply the formalism developed in Section~\ref{sec:formalism} to this example. We model the data as the sum of the true signal and foregrounds multiplied with the antenna gain as before
\begin{eqnarray}
    \bm{v_d} = \left(\matr{I} + \mbf{G}\right) \left(\bm{v_{HI}} + \bm{v_F}\right),
\end{eqnarray}
where the matrix $\mbf{G}$ assigns antenna errors to visibilities in the redundant unstacked space and can be defined as
\begin{equation}\label{equ:G}
    \mbf{G} = \sum_{\nu} \sum_{i = 1}^{2N_a} q_{\nu i} \mbf{\Gamma}_{\nu i}.
\end{equation}
The matrix $\mbf{\Gamma}_{\nu i}$ is the individual error matrix which picks up all the visibilities that involve the $i$-th antenna at the $\nu$-th frequency and assigns the antenna error $q_{\nu i}$ to them. It can be defined as the identity matrix $\mbf{I}$ but only with diagonal elements that correspond to frequency $\nu$ and antenna $i$ being one:
\begin{equation} \label{first_E}
\begin{split}
    (\Gamma_{\nu i})_{(\nu^\prime i^\prime j^\prime)(\nu^{\prime\prime} i^{\prime\prime} j^{\prime\prime})} = &  I_{(\nu^\prime i^\prime j^\prime)(\nu^{\prime\prime} i^{\prime\prime} j^{\prime\prime})} \delta_{\nu \nu^\prime} \delta_{i i^\prime}  \\ & \text{if $i = 1$ to $N_a$},
\end{split}
\end{equation}
and 
\begin{equation} \label{second_E}
\begin{split}
    (\Gamma_{\nu i})_{(\nu^\prime i^\prime j^\prime)(\nu^{\prime\prime} i^{\prime\prime} j^{\prime\prime})} = & I_{(\nu^\prime i^\prime j^\prime)(\nu^{\prime\prime} i^{\prime\prime} j^{\prime\prime})} \delta_{\nu \nu^\prime} \delta_{i j^\prime}  \\ & \text{if $i = N_a+1$ to $2N_a$.}  
\end{split}
\end{equation}
Note that $\delta_{i i^\prime}$ in Eq.~(\ref{first_E}) is changed to $\delta_{i j^\prime}$ in Eq.~(\ref{second_E}) because the $(N_a+1)$-th to $(2N_a)$-th antenna only appear second in a pair.

We now proceed as the previous two examples. Applying the KL filter to data in the redundant unstacked space, we compute the estimated signal $\bm{\hat{v}_{HI}}$ and the estimated foregrounds $\bm{\hat{v}_{F}}$. With the individual error matrix $\mbf{\Gamma}_{\nu i}$ defined in Eq.~(\ref{first_E}) and~(\ref{second_E}), the perturbation estimate Eq.~(\ref{ex3_y_hat_1}) and~(\ref{ex3_y_hat_2}) can now be written as one single equation:
\begin{equation} \label{eq:ex3_perturbation_estimate}
     \hat{y}_{(\nu i)} = \frac{\bm{\hat{v}_{F}}^\dagger(\mbf{\Gamma}_{\nu i}) \bm{\hat{v}_{HI}}}{\bm{\hat{v}_F}^\dagger(\mbf{\Gamma}_{\nu i}) \bm{\hat{v}_F}}.
\end{equation}
This is in fact consistent with the quadratic estimator and normalization operator designed in Eq.~(\ref{eq:E_i_choise}), since Eq.~(\ref{eq:ex3_perturbation_estimate}) is equivalent to having $\matr{E}_{\nu i} = \matr{D}_{\nu i} = \matr{\Gamma}_{\nu i}$.

The perturbation estimate Eq.~(\ref{eq:ex3_perturbation_estimate}) has the same form as the one of the band-pass case [Eq.~(\ref{ex1y_hat_ind})], so the window matrix is identical to Eq.~(\ref{window}) but with all quantities in the redundant unstacked space. We can then recover the gain $q_{\nu i}$ by compensating for the window and finally obtain the cleaned signal by subtracting foreground contamination from the estimated signal in the same way as the previous cases.

\begin{figure}[h!]
\includegraphics[scale=0.34]{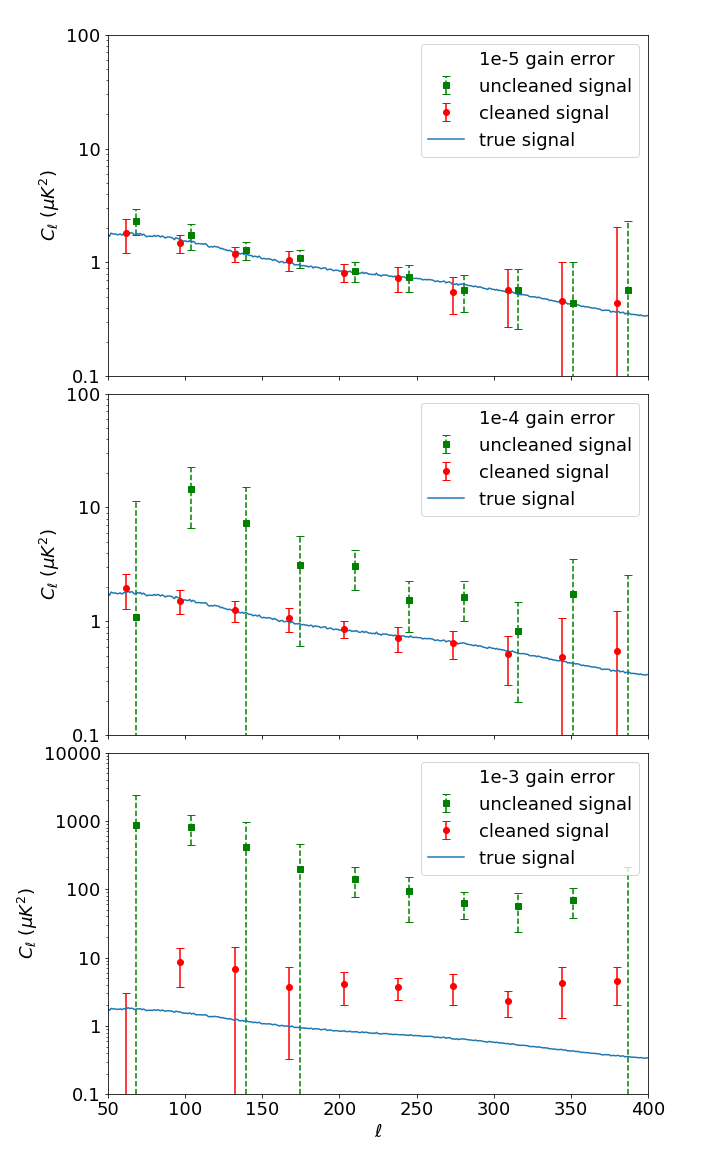}
\caption{\label{third_case_fig} HI power spectrum of the uncleaned, cleaned, and true signal with antenna-dependent error in the unstacked space. Antenna errors have the same set-up as in Fig.~\ref{second_case_fig}. No significant bias is introduced with errors at the $10^{-5}$ level (top panel). At $10^{-4}$ level (middle panel), the foreground residual subtraction removes the bias from the uncleaned signal. At the $10^{-3}$ level (bottom panel), the foreground residual subtraction suppresses foreground bias by two orders of magnitude.}
\end{figure}

We test the foreground residual subtraction algorithm using the identical set-up for antenna errors as in Section~\ref{subsec:antenna_error} and the same sky map as in Section~\ref{subsec:bandpass_error}. Figure ~\ref{third_case_fig} compares HI power of the uncleaned, cleaned, and true HI signal. The results are consistent with the previous two examples: No significant bias is introduced by antenna errors at the $10^{-5}$ level, but with larger antenna errors, the bias in the uncleaned signal becomes larger. The foreground residual subtraction algorithm is able to remove the bias when antenna errors are at the $10^{-4}$ level and suppress the bias by two orders of magnitude with errors at the $10^{-3}$ level.

\section{Discussion}
\label{sec:discussion}

\subsection{Estimator form}

We have developed an algorithm that estimates telescope systematic errors from linearly filtered data with a quadratic estimator, and then subtracts the systemics-induced foreground contamination. The algorithm is motivated by the fact that the estimated signal is dominated by foreground residuals which correlate with the foregrounds. Therefore, cross-correlation between the contaminated signal channels and estimated foregrounds isolates the systematics. 

This form, where we cross-correlate the estimated signal and foregrounds to estimate the foreground residuals, has a number of qualitative advantages. First, it explicitly targets precisely the thing we wish to eliminate: foregrounds leaking into the signal channel. The fact that the foregrounds themselves are typically measured at very high signal-to-noise ratio means we have an essentially noiseless template with which to draw out the residual foregrounds. Thus spurious correlations from either noise or 21-cm signal in the signal channel will be sub-dominant.

The cross-correlation also provides a means to control the non-linearity of the method. Other non-linear methods result in significant and hard-to-characterize signal loss. Here, it is only signal that spuriously correlates with the foregrounds that can be lost. The foregrounds themselves have few degrees of freedom compared to the signal due to their spectral smoothness, and estimates of the foregrounds have little signal contamination due to the difference in their brightness. Our algorithm introduces additional degrees of freedom in the number of ways we cross-correlate the data (\textsl{i.e.},~the number of systematics-related parameters to be estimated), which must be substantially smaller than the number of degrees of freedom in the data itself. However, in order for signal to be lost in the foreground residual cleaning process, the power in the signal channel must correlate with well-determined foregrounds. This mitigates over-subtraction and signal loss, even when the number of free parameters is relatively large. Indeed a perturbative expansion in the small parameters describing the systematics, and in the ratio of signal to foreground powers, provides control over how much foregrounds are not removed by our algorithm. A similar expansion can be used to estimate signal loss from the non-linearity, although we have not performed this calculation.

Having motivated a method that cross-correlates the signal and foreground estimates, what remains is to determine \emph{how} to cross-correlate them. That is, what set of transformations $\matr{E}_i^\dagger$ should be applied to the foreground estimate prior to cross-correlation to draw out the residual contamination? In this work, we assumed the systematics are described by a parametric model with unknown (but small) parameters $g_i$ and use a quadratic estimator framework to determine the cross-correlation that returns the parameters. This formalism applies to errors in the signal-chain gains, and should be straight-forward to generalize other systematics with limited numbers of degrees of freedom as discussed below.

For some types of systematics, such as complex variations in the primary beam, it may not be practical to write down a parameterized model. Then, it would not be possible to construct a quadratic estimator, since there is no parameter to estimate. However, the basic idea of cross-correlating the signal estimate with the foreground estimate, in order to draw out residual foregrounds, could still be valid. One would need to determine---perhaps empirically---through what transformations and with what symmetries the foregrounds are leaking into the signal, such that the subspace over which to cross-correlate can be determined. While we believe this generalization is very promising, we leave further consideration to future work.

In our analysis, we have chosen the quadratic estimator $\matr{E}_i$ to be the perturbation base matrix $\matr{\Gamma}_i$ based on the intuition that in order to estimate a particular perturbation $g_i$, we want to cross-correlate all the data points corrupted by it while leaving out those that are unaffected. We verified this choice with simulations in Section~\ref{sec:examples} by showing that the algorithm reduces foreground contamination in the power spectrum of the cleaned signal by one to three orders of magnitude.

One particular characteristic of our formalism developed in Section~\ref{sec:formalism} is that, in order to estimate systematics, the quadratic estimator can be chosen such that it is independent of any signal or foreground model (for example, when we choose $\matr{E}_i = \matr{\Gamma}_i$). Then, the only model-dependent components in our systematics estimation are the window matrix $\matr{W}$ and linear foreground filter $\matr{K}$. However, the window matrix can, in fact, be approximated using data only instead of using assumed foreground and signal models. This is because although the window matrix requires the foreground covariance, our data is already an excellent measurement of the foregrounds (given that $\vect{s} \ll \vect{f}$ and $g_i \ll 1$) and can be used to approximate the covariance . 

We can approximate the denominator of the window matrix, Eq.~(\ref{eq:formalism_W}), as
\begin{equation}
    \text{Tr}(\matr{A}^\dagger \matr{D}_i \matr{A} \matr{F}) \approx \vect{\hat{f}}^\dagger \matr{D}_i \vect{\hat{f}},
\end{equation}
and approximate the numerator as
\begin{equation}
    \text{Tr}(\matr{E}_i\matr{K} \mbf{\Gamma}_{i^\prime}\mbf{F}\matr{A}^\dagger) \approx \vect{\hat{f}}^\dagger \matr{E}_i\matr{K} \mbf{\Gamma}_{i^\prime} \vect{d}.
\end{equation}
So the expression for the window matrix
\begin{equation}
    W_{i i^\prime} \approx \frac{\vect{\hat{f}}^\dagger \matr{E}_i\matr{K} \mbf{\Gamma}_{i^\prime} \vect{d}}{\vect{\hat{f}}^\dagger \matr{D}_i \vect{\hat{f}}}
\end{equation}
does not explicitly depend on signal or foreground models, although the linear filter $\matr{K}$ does. In our analysis, we choose the KL filter as the linear foreground filter $\matr{K}$, but the formalism is, in fact, independent of this choice. Therefore, any other choice for $\matr{K}$ will work as well.

\subsection{Limiting factors on current results}
\label{subse:limiting_factors}

Section~\ref{sec:examples} shows that our foreground residual subtraction algorithm can---in the context of our somewhat simple simulations---successfully remove foreground contamination with band-pass errors up to the $10^{-3}$ level and antenna-dependent errors up to the $10^{-4}$ level. Several factors prevent the algorithm from achieving better results with larger errors: most notably the second order terms of perturbations, statistical noise, and instrumental noise. 

Derivations in Section~\ref{sec:formalism} ignored all terms that involved the square and higher powers of perturbations. Particularly, we dropped the second order terms from the data covariance matrix in Eq.~(\ref{eq:data_cov}) and from the window matrix in Eq.~(\ref{eq:another_way}). As a result, we can subtract foreground residuals only up to the first order in amplitude of perturbations as shown by Eq.~(\ref{v_HI_tilde}). This naturally leaves foregrounds at the order of $g^2 v_F$, which can be neglected as long as $g^2 v_F \ll v_{HI}$. Panel (a) and (b) of Fig.~\ref{visibilities} show that the foreground visibilities are almost $10^5$ times brighter than HI in the simulations. This suggests that if band-pass errors are at the $10^{-3}$ level or below, their second order terms can be safely ignored. This is consistent with the results shown in Fig.~\ref{first_case_fig}. In comparison, Fig. ~\ref{second_case_fig} and~\ref{third_case_fig} show that the foreground residual subtraction does not recover the theoretical HI power spectrum well with $10^{-3}$-level antenna-dependent errors. This is because the rms value of the antenna-dependent errors is a few times larger than that of the band-pass errors.

Statistical noise can also introduce bias to the cleaned signal. In Eq.~(\ref{eq:recovered_gain}), we compensated the window by applying the pseudo-inverse of the window matrix to the estimated perturbations $\hat{y}$. Strictly speaking, we should compensate the window for the ensemble mean of the the perturbation estimate $\langle\hat{y}\rangle$, but in reality we only have one sky to observe. The difference between $\hat{y}$ and $\langle\hat{y}\rangle$ results in an error in the recovered perturbations $\hat{g}$. They are subsequently used to subtract foreground contamination at the linear order and thus leave foreground residuals at the order of $\left(\langle\hat{y}\rangle - \hat{y}\right) v_F$. Figure~\ref{gain} shows that the perturbation estimates (yellow square data points) and their ensemble averages (green triangle data points) have percent-level differences. This implies that statistical noise does not introduce significant bias with perturbations at the $10^{-3}$ level or below since the residual should be smaller than the HI signal at that level. This is again consistent with the results shown in Section~\ref{sec:examples}.

Lastly, instrumental noise can also affect foreground residual cleaning. The perturbation estimate relies on the estimated signal being dominated by foreground contamination at the order of $g v_F$. However, in the case of relatively small perturbations and large instrument noise, the later may dominate the former. Even though noise is removed during power spectrum estimate by cross-correlating data from two seasons, a dominating noise term in the estimated signal causes the perturbation quadratic estimator to pick up noise instead of the gain errors, which subsequently affects foreground residual subtraction. In Section~\ref{subsec:bandpass_error} and~\ref{subsec:antenna_error_unstacked}, the sky map was observed for 120 days such that the resulting noise is about a few times brighter than the HI signal but more than 1000 times weaker than the foreground. However, in Section~\ref{subsec:antenna_error}, the integration time of each sky map is decreased to 8 days, so the noise becomes larger. As a result, cleaned signals shown in Fig.~\ref{second_case_fig} has bigger error and higher bias than the ones in Fig.~\ref{first_case_fig} and~\ref{third_case_fig}.

\subsection{Comparison with gain calibration literature}
\label{subsec:gaincal_literature}

In order to remove foreground residuals from linearly filtered data, estimating gain perturbations is an essential step of the algorithm. Traditional gain calibration methods can be summarized into two categories: the ``sky-based" calibration and the ``redundant calibration" \citep{Byrne_2021}. The former uses sky and instrument models to produce simulated data and compares them to the real data to estimate the gains, while the later constrains antenna gains by checking consistencies between redundant baseline measurements. Both methods are Bayesian inference of complex gain parameters from the data and the model of the data. In other words, they aim to find the gain parameters $\bm{g}$ that maximize the posterior probability $P(\bm{g}|\bm{v})$ from the visibility data $\bm{v}$.  

Our gain estimation, in comparison, uses a quadratic estimator, $\mbf{E}_i$ in equation (\ref{eq:pert_estimate}), to estimate the gains by cross-correlating two different sets of linearly filtered data. It is essentially a maximum likelihood approach, which estimates the gain parameters $\bm{g}$ that maximize the likelihood function $\mathcal{L}(\bm{g}|\bm{\hat{f}}, \bm{\hat{s}})$, where $\bm{\hat{f}}$ and $\bm{\hat{s}}$ are the estimated foreground and signal, respectively. The estimated foreground correlates with the estimated signal precisely due to foreground residuals in the estimated signal. Compared with the traditional methods, the special trait of our approach is that it directly targets what we want to remove: the foreground residuals. 

Our gain estimate approach can enhance the traditional gain calibration methods. First, both sky-based calibration and redundant calibration rely on accurate sky models. In sky-based calibration, sky models are needed to produce simulated data. Redundant calibration can in principle estimate the relative gain among antennae without the knowledge of a sky model, but to break gain parameter degeneracies and produce physical calibration results, absolute calibration must be done with the assumption of a sky model \citep{10.1093/mnras/stu1773, Byrne_2019, Kern_2020}. In reality, sky models contain inaccurate intensities for known sources and can have other faint sources missing. These inaccuracies lead to errors in gain calibrations \citep{10.1093/mnras/stu268, Barry_2016, 10.1093/mnras/stw2277, Ewall-Wice_2016, Joseph_2020}. In comparison, our approach of gain estimation does not reply on a specific sky model. Instead, it requires the covariance matrices of the foreground and signal to train the linear filter. Our approach is thus immune from inaccurate and incomplete prior knowledge of the sky and only needs statistical information of the foregrounds and signal.

In addition, sky-based calibration relies on an accurate understanding of instrument response in order to map the sky into visibilities. Errors in instrument response, such as baseline perturbations, can thus produce errors in modelled visibilities which can propagate through the calibration process \citep{Barry_2016, Ewall-Wice_2016}. Likewise, baseline perturbations can cause non-redunancies in baselines that would otherwise observe identical sky information. This breaks the basic assumption of redundant calibration and leads to calibration error \citep{Joseph_2018, Li_2018, Orosz_2019}. Extensions to traditional calibration methods can account for instrument response error by modelling beam response via either direct measurement or simulation software \citep{Sullivan_2012, Lanman_2019, Mort_2010, Jagannathan_2017}. However, such extensions require additional measurement or modelling prior to gain calibration. 

In comparison, estimating instrument response errors can be included in our gain estimation algorithm. Not only can we apply our algorithm to estimate antenna gains but also to any perturbation that can be parametrized. For any set of perturbation parameters $\lambda_\alpha$, if we can describe their effect on the visibilities as 
\begin{equation} \label{extension}
    \bm{v}_{ij} = \bm{\bar{v}}_{ij} + \sum_\alpha \lambda_\alpha \frac{\partial \bm{v}_{ij}}{\partial \lambda_\alpha},
\end{equation} 
where $\bm{\bar{v}}_{ij}$ is the true visibility, then we can estimate $\lambda_\alpha$ in the same way that we estimate antenna gains. We will discuss this idea in more details in Section \ref{subsec:other_systematics}.

\subsection{Comparison with foreground removal literature}
\label{subsec:foreground_literature}

Our foreground removal technique combines a traditional linear filter with a nonlinear quadratic estimator to estimate the gain errors and clean foreground residuals. We used the KL filter \cite{COBE_KL} as the linear filter in our examples, but other linear filters, such as the delay filter \cite{delay_filter}, can work with the algorithm as well. Our findings in Section~\ref{sec:examples} suggest that using the KL filter alone results in bright foreground residuals. This is because telescope systematics introduce non-smooth spectral features to foregrounds, thus violating the basic assumption of traditional linear filters that the foregrounds are spectrally smooth. Compared with using a traditional linear filter alone, our hybrid foreground filtering technique can suppress foreground residuals for one to three orders of magnitude, as shown in Fig.~\ref{first_case_fig},~\ref{second_case_fig}, and~\ref{third_case_fig}. 

Our hybrid method has advantages over traditional non-linear foreground removal techniques as well in some aspects. Traditional non-linear foreground removal methods, such as the principal component analysis and its related non-parametric component separation algorithms \citep{PhysRevD.83.103006, abf+15, bmb+16, PCA_cite, wab+14, Zhang_2016, Cunnington_2019, ord16, Carucci_2020}, project out the brightest modes from the total data covariance with the assumption that the brightest modes are mostly dominated by foregrounds. Such techniques are robust to some systematics because the brightest modes are discarded regardless of whether they are spectrally smooth or not. However, brightest modes can also contain significant amount of signal, so oversubtraction could become an issue that hinders signal detection \citep{Switzer_2015}. Moreover, zeroing out the brightest modes cannot address the possibility of foreground leakage into the less bright modes. 


In comparison, traditional linear filters which are part of our hybrid algorithm target the smooth component of the data which is dominated by foregrounds. This helps one control how much signal is lost and estimate how much foreground is left in the remaining modes \citep{Shaw_2014}. Therefore, the hybrid foreground removal technique essentially combines the advantages of both the linear and non-linear filters, meaning it is easy to characterize signal loss and at the same time more robust to systematics.

\subsection{Extension to other systematics}
\label{subsec:other_systematics}

Our algorithm can potentially be applied to subtract other types of systematics in a generic radio interferometry telescope. The essence of estimating systematics is to calculate the pertubration base matrix $\mathbf{\Gamma}$ as in Eq.~(\ref{eq:formalism_pert_matrix}). Generically, Eq.~(\ref{eq:formalism_pert_matrix}) projects each systematics in the form of a coefficient $g$ and a derivative matrix $\mathbf{\Gamma}$ that characterizes the system response to a particular systematics. Depending on the specific types of systematics,   $\mathbf{\Gamma}$ can be a function of different instrumental parameters such as frequency, antenna, or baseline distance, etc. For example, for baseline distortions induced by feed position shifts, one can parameterize the derivative matrix for each antenna pair as a function of the perturbations on the baseline distance $\mathbf{u}$ so that $\mathbf{\Gamma}\propto(\Delta u_i+ \Delta u_j)$, with $i$ and $j$ being the coordinates of the two feeds.  

The main limitation of our approach is that we need to be able to write down a parameterized model for the systematics with derivative matrices. This is convenient to do for gain and baseline perturbations, but for systematics like beam perturbations, it may be unrealistic to calculate the derivative matrix simply with an analytical parameterization. In this case, one may vary relevant instrumental parameters through simulation, and numerically calculate the derivative matrix. For example, one can vary the primary beam width or the beam pointing angle to quantify the systematics induced by those factors and subsequently subtract them out using our algorithm. Nevertheless, our algorithm is potentially  applicable to a variety of systematics as long as the characteristic derivative matrix can be deduced. 

\section{Conclusions}
\label{sec:conclusions}

In this paper, we have developed a novel hybrid foreground removing algorithm for 21-cm intensity mapping experiments by combining a traditional linear filter with a non-linear quadratic estimator. With simulations of a small-scale compact array, we have demonstrated that we can suppress foreground residuals in the linearly filtered 21-cm signal by one order of magnitude when antenna-dependent complex gain errors are at the level of $10^{-4}$ and nearly two orders of magnitude at the level of $10^{-3}$. In the case of $10^{-4}$-level errors, the signal after foreground residual cleaning recovers the theoretical HI power spectrum. 

Compared with traditional linear methods, the hybrid algorithm is more robust to systematics by calibrating them using a quadratic estimator and subsequently subtracting the induced foreground contamination from the data. Compared with traditional non-linear methods, the hybrid algorithm is easier to characterize and quantify signal loss, due to perturbative control over non-linearities. Our method thus combines the advantages of both linear and non-linear methods while each compensates for the other's drawbacks.

At the same time, there is room for improvement on the current version of the hybrid algorithm. At the end of Section~\ref{sec:formalism}, we picked a simple form for the quadratic estimator and applied it to three examples in Section~\ref{sec:examples} to intuitively illustrate the idea of the technique. However, we could instead use the optimal quadratic estimator to further reduce uncertainties on the estimated calibration parameters. Moreover, due to computational limitations, we simulated the data from a small $5\times5$ square array. A larger array of a realistic size will have many more baselines and therefore incorporate more information from the sky, which will decrease the statistical noise of our estimated systematics and further improve our results. Lastly, it is worth mentioning that in the current version of our algorithm, we only compute and remove the foreground residuals that come from first order terms in amplitude of the perturbations. In principle, the second order terms can be computed and removed from the linearly filtered data as well. These improvement will be explored in future studies. 

While our simulations are simplistic, they nonetheless capture the essential features that make foreground removal difficult and thus demonstrate the potential of our algorithm.
Precisely how effective the algorithm is could change somewhat with more realistic simulations, as well as the design of the instrument and survey.
The telescope systematics used in the simulations are limited to band-pass errors and antenna-dependent complex gain errors. However, as mentioned in Section~\ref{sec:discussion}, any type of error that can be parameterized as in equation (\ref{extension}) can be estimated by our algorithm. Thus, a promising future direction is to generalize our algorithm to calibrate other types of systematic error, such as baseline errors and beam errors. In addition, the simulation pipeline we use to test the hybrid algorithm does not include sky polarization, but studies have shown that linear filters, such as the KL filter, can be applied to polarized data as well \citep{m_mode_richard}. Applying our hybrid algorithm to polarized data is another aspect to be explored in future studies. 

Clearly, there will be many opportunities to generalize our hybrid algorithm for potentially a wide range of low frequency intensity mapping experiments. With the work presented in this paper, we hope that our hybrid algorithm can provide a new powerful tool to mitigate the foreground contamination problem.

\begin{acknowledgements} 
We would like to thank the CHIME collaboration for their feedback on this paper and especially Simon Foreman and Richard Shaw for their suggestions. We would also like to thank the intensity mapping group in the University of Manchester for providing the realistic simulation pipeline to generate the test dataset in our work. 
This study is funded by an NSF grant (2008031).
\end{acknowledgements}

\appendix
\section{DETAILED ANALYSIS OF THE TOY EXAMPLE}
\label{app:toy_model}

In order to demonstrate that the cleaning algorithm
indeed removes the effect of the foregrounds for the
toy model, we will start by re-writing 
Eq.~(\ref{eq:f_s_hat_pert}) as

\begin{eqnarray}
\label{app_eq:f_s_hat_pert}
\begin{split}
   \hat{f}_p =& \alpha f_p + \frac{1}{N} \sum_{\nu^\prime} 
   m_{\nu^\prime p}, \\
   \hat{s}_{\nu p} =&  m_{\nu p}
       - \frac{1}{N} \sum_{\nu^\prime} 
       m_{\nu^\prime p} + 
        \beta_\nu f_p
\end{split}
\end{eqnarray}

\noindent where

\begin{eqnarray}
\label{app_eq:m_alpha_beta}
\begin{split}
   m_{\nu p} &= s_{\nu p} \left( 1 + g_{\nu} \right), \hspace{0.25in}
   \alpha =  1 + \frac{1}{N}\sum_{\nu} g_{\nu}, \\
   \beta_\nu &= g_\nu - \frac{1}{N}\sum_{\nu^\prime} g_{\nu^\prime}.
\end{split}
\end{eqnarray}

By plugging Eq.~(\ref{app_eq:f_s_hat_pert})
and (\ref{app_eq:m_alpha_beta}) into the definition
of $\hat{y}_\nu$ in Eq.~(\ref{eq:y_hat}) and re-arranging
(including the Taylor-expansion of the denominator)
we find

\begin{eqnarray}
\label{app_eq:y_hat}
\begin{split}
   \hat{y}_{\nu} =& \left\{\frac{\beta_\nu}{\alpha} + 
   \frac{1}{\alpha \sum_p f_p^2} \left[ \sum_p f_p m_{\nu p} + 
   \right. \right. \\
   &\left. \left. \frac{1}{N} \sum_{\nu^\prime p} 
   \left[ f_p m_{\nu^\prime p} \left( \frac{\beta_\nu}{\alpha} -1 \right)
   + \frac{1}{\alpha} m_{\nu p} m_{\nu^\prime p}\right] - 
   \right. \right. \\
   & \left. \left. \frac{1}{\alpha N^2} \sum_{\nu^\prime \nu^{\prime\prime} p} 
   m_{\nu^\prime p} m_{\nu^{\prime\prime} p} \right] \right\} \times \\
   & \left\{1- \frac{1}{\alpha \sum_p f_p^2} \left[ 
   \frac{2}{N} \sum_{\nu^\prime p} f_p m_{\nu^\prime p}+ 
   \right. \right. \\
   & \left. \left. \frac{1}{\alpha N^2} \sum_{\nu^\prime \nu^{\prime\prime} p} 
   m_{\nu^\prime p} m_{\nu^{\prime\prime} p} \right] + \cdots \right\}.
\end{split}
\end{eqnarray}

We are interested in $ \langle \hat{y}_\nu \rangle$, 
so we need to calculate the ensemble
average of Eq.~(\ref{app_eq:y_hat}) over signal and foreground 
realizations while keeping the $g_\nu$ fixed. 

The leading term in $\hat{y}_\nu$ is $\beta_\nu/\alpha$ 
[the first term within the first curly brackets in 
Eq.~(\ref{app_eq:y_hat})], which
is a function of $g_\nu$ only. 
Terms of the form $\sum_p f_p m_{\nu p}$ are zero on average,
so the next term in importance in the expansion of 
$ \langle \hat{y}_\nu \rangle$ has the form
$\langle n/d \rangle $ where

\begin{eqnarray}
\label{app_eq:n_d}
\begin{split}
    n =& \frac{1}{\alpha^2 N}\left( \sum_{\nu^\prime p} m_{\nu p} m_{\nu^\prime p}
        - \frac{1}{N} \sum_{\nu^\prime \nu^{\prime\prime} p} 
         m_{\nu^\prime p} m_{\nu^{\prime\prime} p}\right) \\
    d =& \sum_p f_p^2.
\end{split}
\end{eqnarray}

This term can be approximated as\footnote{Equation~\ref{app_eq:n_d_exp}
is the first term of the expectation of the function 
$h(n, d)=n/d$ Taylor expanded about
$\left( \langle n \rangle, \langle d \rangle \right)$. The next term
in the expansion is a factor of $\sim M$ smaller, 
so it can be safely neglected.}

\begin{eqnarray}
\label{app_eq:n_d_exp}
\begin{split}
   \left \langle \frac{n}{d} \right \rangle \approx& 
   \frac{\langle n \rangle}{\langle d \rangle} \\
   =& \frac{\sigma_s^2}{\alpha^2 N \sigma_f^2} 
   \left [\left(1+g_\nu\right)^2 - \frac{1}{N}\sum_{\nu^\prime} 
   \left(1+g_{\nu^\prime}\right)^2 \right].
\end{split}
\end{eqnarray}

From Eq.~(\ref{app_eq:n_d_exp}), $\langle n/d \rangle $ is of order 
$\sim \sigma_s^2 \sigma_g/\left(N \sigma_f^2\right)$ where $\sigma_g\ll 1$
is the scale of the gain perturbations. This term is
negligible compared to $\beta_\nu/\alpha$ 
(which goes roughly as $\sim \sigma_g$)
since $N \gg 1$ and $\sigma_s^2/\sigma_f^2\ll 1$. Thus

\begin{equation}
\label{app_eq:y_hat_mean}
\begin{split}
    \langle \hat{y}_\nu \rangle =& \frac{\beta_\nu}{\alpha} + \mathcal{O}\left(\frac{\sigma_s^2 \sigma_g}{N\sigma_f^2}\right) 
\end{split}             
\end{equation}

\noindent which is the result shown in Equation~\ref{eq:y_hat_mean}.

The `cleaned' signal is defined in Equation~\ref{eq:s_tilde} as
$\tilde{s}_{\nu p}=\hat{s}_{\nu p} - \hat{y}_\nu \hat{f}_p$.
From Equations~\ref{app_eq:f_s_hat_pert} and \ref{app_eq:y_hat},
$\tilde{s}_{\nu p}$ is indeed free of the term $\beta_\nu f_p$ which is
the source of foreground contamination in $\hat{s}_{\nu p}$
(Equation~\ref{app_eq:f_s_hat_pert}) and in its variance 
(Equation~\ref{eq:s_hat_var}). However, it still has
residual foreground contamination due to the higher order terms 
(beyond $\beta_\nu/\alpha$) in the
expansion of $\hat{y}_\nu$.
An inspection of these terms reveals that they are zero on average
and that their contribution to the variance 
of $\tilde{s}_{\nu p}$ is below $\sim \sigma_s^2/M$,
which for $M \gg 1$ are too small compared to 
$\left\langle m_{\nu p}^2\right\rangle \approx \sigma_s^2 \left(1+2g_\nu\right)$,
the dominant term in the variance of the cleaned signal. 
The final expression for in 
$\left\langle \tilde{s}_{\nu p}^2\right\rangle$ is given in 
Equation~\ref{eq:s_tilde_var}, where $\mathcal{O}\left(\sigma_g, 1/M\right)$
means that the next terms in the expansion are of order $\sigma_g$ and $1/M$.
On average, the cleaned signal is free of foreground bias to all orders.

\section{DERIVATION OF THE WINDOW MATRIX}
\label{sec:formalism_app}
The way we calculate the window matrix is to approximate the right-hand side of Eq.~(\ref{start_to_compute_W}) by taking the ensemble average of the numerator and the denominator and finding their ratio. Namely, we have
\begin{equation}
    \langle \hat{y}_i \rangle \approx  \frac{ \langle \vect{\hat{f}}^\dagger \matr{E}_i \hat{\vect{s}} \rangle}{\langle \vect{\hat{f}}^\dagger \matr{D}_i \vect{\hat{f}} \rangle}.
\end{equation}
To see why this approximation is reasonable, we can rewrite Eq.~(\ref{eq:pert_estimate}) as $\vect{\hat{f}}^\dagger \matr{D}_i \vect{\hat{f}} \hat{y}_i = \vect{\hat{f}}^\dagger \matr{E}_i \hat{\vect{s}}$ (with $b_i = 0$), and then take the expectation to get
\begin{equation} \label{eq:another_way}
    \langle \vect{\hat{f}}^\dagger \matr{D}_i \vect{\hat{f}} \hat{y}_i \rangle = \langle \vect{\hat{f}}^\dagger \matr{E}_i \hat{\vect{s}} \rangle.
\end{equation}
Note that on the left-hand side of Eq.~(\ref{eq:another_way}), the perturbation estimate $\hat{y}_i$ is already at the order of $g_i$. Since we are interested in the linear order of perturbations only, we can drop the perturbation term [the term that involves $\matr{G}$ in the first line of Eq.~(\ref{eq:form_f_hat})] in $\vect{\hat{f}}$ on the left hand side. Eq.~(\ref{eq:another_way}) now becomes
\begin{equation} \label{eq:another_way_linear}
    \langle (\vect{s} + \vect{f})^\dagger \matr{A}^\dagger \matr{D}_i \matr{A}(\vect{s} + \vect{f})  \hat{y}_i \rangle = \langle \vect{\hat{f}}^\dagger \matr{E}_i \hat{\vect{s}} \rangle.
\end{equation}
Suppose the choice of the quadratic estimator $\matr{E}_i$ and the normalization operator $\matr{D}_i$ is good such that $\hat{y}_i \approx g_i$, then $\hat{y}_i$ is only weakly dependent on the sky signal $\vect{s}$ and $\vect{f}$ under the assumption that errors of the instrument $g_i$'s are independent of the sky signal. 
This was the case for the toy example described in Section~\ref{sec:toy_example} and Appendix~\ref{app:toy_model}, where
we analytically showed that higher order terms in the expansion of
$y$ are orders of magnitude smaller than the gain perturbation terms
on average.
Therefore, we can separate the expectation value of $\hat{y}_i$ from the rest of the left-hand side and write Eq.~(\ref{eq:another_way_linear}) as
\begin{equation} \label{eq:another_way_linear_2}
    \langle (\vect{s} + \vect{f})^\dagger \matr{A}^\dagger \matr{D}_i \matr{A}(\vect{s} + \vect{f})\rangle  \langle\hat{y}_i \rangle = \langle \vect{\hat{f}}^\dagger \matr{E}_i \hat{\vect{s}} \rangle.
\end{equation}
Simplify both sides of Eq.~(\ref{eq:another_way_linear_2}), we get
\begin{equation} \label{eq:another_way_linear_3}
    \text{Tr}[\matr{A}^\dagger \matr{D}_i \matr{A} (\matr{S} + \matr{F})] \langle\hat{y}_i \rangle = \text{Tr}(\matr{E}_i \matr{C}^{sf}).
\end{equation}
Replacing the covariance $\matr{C}^{sf}$ with Eq.~(\ref{eq:sf_cov}), and divide Eq.~(\ref{eq:another_way_linear_3}) by the normalization factor $\text{Tr}[\matr{A}^\dagger \matr{D}_i \matr{A} (\matr{S} + \matr{F})]$ from the left-hand side, we finally get Eq.~(\ref{eq:window_result}) which we simply stated in Section~\ref{sec:formalism}.

\section{SIMULATION PIPELINE DETAILS}
\label{app:pipeline}

\begin{figure*}
  \includegraphics[width = 0.98\hsize]{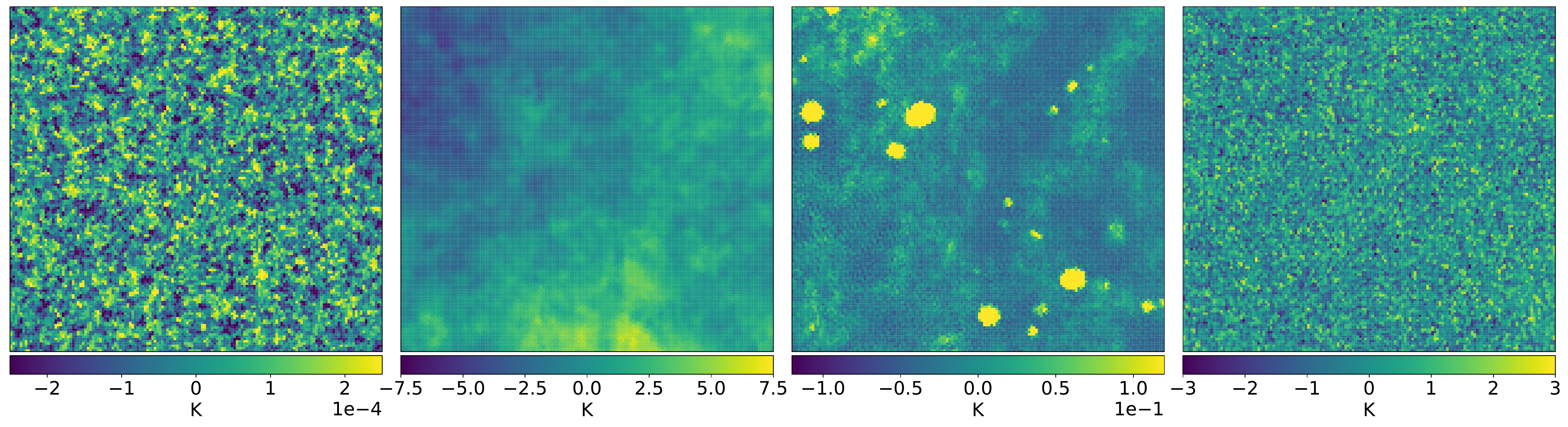}
  \includegraphics[width = 0.98\hsize]{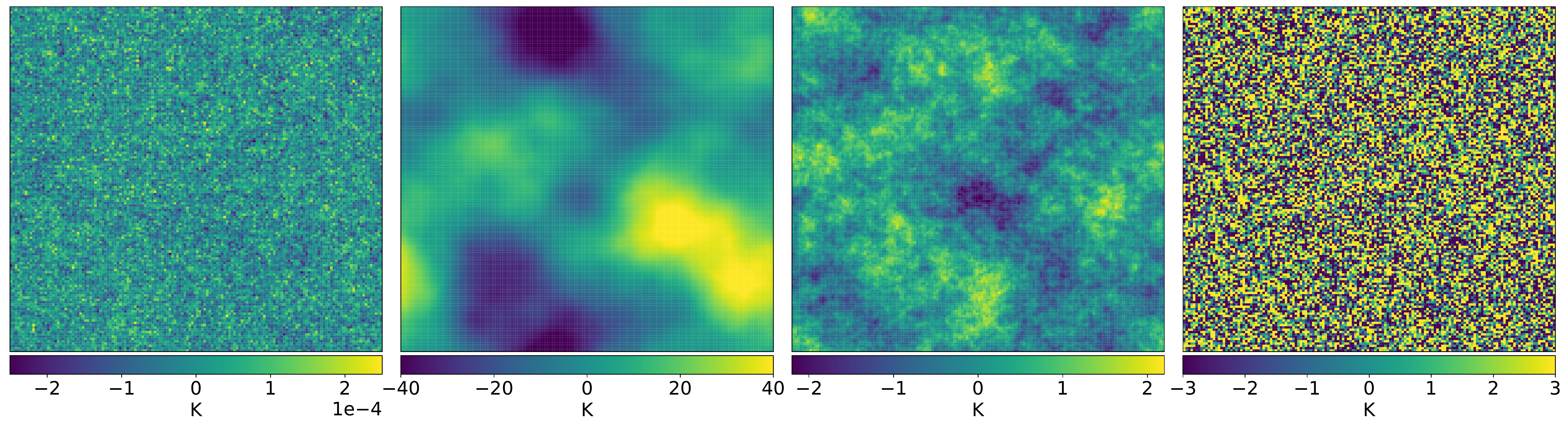}
\caption{Upper panels: The test dataset generated using  the realistic simulation pipeline based on \protect\cite{hdb+18} and \protect\cite{odb+18}. From left to right, we show a map of the HI, synchrotron, free-free and point source at a random location in the sky. Lower panels: The simulated maps using a simplified approach to compute  prior covariance matrices for the foreground filter. Each panel shows a  $30^{\circ}\times30^{\circ}$ patch on the sky with  $150\times150$ pixels at   the first frequency channel of 400\,MHz.  }
\label{fig:Tmaps}
\end{figure*}
\subsection{Sky Model}\label{sec:sky}

As described in Section~\ref{sec:formalism}, the KL-based foreground filter requires prior knowledge of the sky components encoded in the covariance matrices of the signal $\matr{S}$ and foregrounds $\matr{F}$. For simplicity, we compute the prior covariance matrices from simulated Monte Carlo (MC) realizations given a simple angular power spectrum and frequency dependency of each component.  However, we adopt an independent and  more realistic sky model based on  \cite{hdb+18} and \cite{odb+18} to generate input maps as our test dataset in order to verify the foreground removal algorithms. This is to simulate the scenario  that as long as our prior knowledge of the sky components is statistically correct, our foreground filter is insensitive to the exact model mismatch between the prior knowledge and the test data. Both the simplified prior simulations and the more realistic test data sets are described in this section. 

\subsubsection{HI emission}\label{sec:HIsim}
The mean brightness temperature of the HI signal as a function of redshift is computed following \cite{bbd+13} by
\begin{equation}
\bar{T}_{\rm obs}(z) = 44\mu\mathrm{K}\left(\frac{\Omega_{\rm HI}h}{2.45\time10^{-4}}\right)\frac{(1+z)^2}{E(z)}\,,
\end{equation}
where $\Omega_{\rm HI}$ is the neutral HI fraction  assumed to be constant over redshift at $\Omega_{\rm HI} = 6.2\times10^{-4}$, $h = H_0/100$\,km\,s$^{-1}$\,Mpc$^{-1}$ and $E(z) = H(z)/H_0$ describing the Hubble expansion.

The HI angular power spectrum is computed by using the Limber approximation \citep{limber53}, which is a good approximation to $\ell\gtrsim50$ assuming a flat-sky,
\begin{equation}
  C_{\ell} = \frac{H_0b_{\rm HI}^2}{c}\int dz E(z)\left[\frac{\bar{T}_{\rm obs}(z)D(z)}{r(z)}\right]^2P_{\rm cdm}\left(\frac{\ell + 0.5}{r}\right)\,,
\end{equation}
where $b_{\rm HI}$ is the HI bias assumed to be constant at unity for simplicity in our simulation, $r(z)$ is the comoving distance out to redshift $z$, $D(z)$ is the growth factor, and $P_{\rm cdm}$ is the cold dark matter power spectrum computed using the \textsc{camb} software \citep{lb02}.

We simulate 50 equally-spaced frequency channels between 400 and 500 MHz.  At each frequency, a \textsc{healpix} HI map realization with \textsc{nside = 256} is generated given its angular power spectrum using the \textsc{synfast} module provided by the \textsc{healpix} package \citep{ghb+05}. Although the simulated HI map realizations are full sky maps, in order to minimise the strong  emission from the other foreground components in the Galactic plane, we mask out the Galactic plane by using the \emph{Planck} 2015 Galactic plane mask with $80\%$ unmasked sky \footnote{Planck Legacy Archive: http://pla.esac.esa.int}.  We arbitrarily select 10 separated sky locations outside of the mask, and a $30^{\circ}\times30^{\circ}$ sky patch centred at each selected location is projected into a 2D Cartesian patch with a pixel size of $150\times150$  as our input  HI dataset. The upper left panel in Fig.\,\ref{fig:Tmaps} shows one example of the HI patch at the first frequency of 400\,MHz.

For the prior covariance matrix to construct  the foreground filter, we adopt  a simpler simulation while preserving the statistic properties.  We create a 2D field of Gaussian distributed random complex numbers with a mean of 0 and  a standard deviation of 1. We compute the real Fourier frequency for each side as
\begin{equation}
\ell_{x,y} = \frac{k}{W_{\rm size}/2\pi}\;, k = \left[1, ..., \frac{N_{\rm pix}}{2-1}, \frac{N_{\rm pix}}{2}\right]\,,
\end{equation}
where $W_{\rm size}  = 30^{\circ}$ is the size of the patch per side converted into radian.  $2\pi$ is to convert the Fourier frequency into the unit of angular scale, i.e., multipole. $N_{\rm pix} = 150$ is the number of pixels per side. The radial magnitude of the  2D Fourier frequency vector is then $\ell_{\rm mag}  = \sqrt{\ell_x^2 + \ell_y^2}$. We define the scale-dependent power spectrum in the form of 
\begin{equation}\label{equ:simps}
P_{\ell} = A\left(\frac{\ell_{\rm mag}}{\ell_{\rm ref}}\right)^{\alpha}\,,
\end{equation}
where $\ell_{\rm ref} = 200$ is the reference scale at which the power spectrum has an amplitude of $A$. $\alpha$ is the power spectrum scale factor  to scale the power with respect to the angular scale.  Depending on the sky component, we choose different values of $A$ and $\alpha$ so that the simulated patches are in the same order of magnitude as the realistic input dataset. The value of $\alpha$ is selected to preserve the  scale-dependent morphology for each  component. For example, the small scale HI signal has a smaller $\alpha$ value compared with the synchrotron emission which is diffused over large scales. We choose $A_{\rm HI} = 10^{-13}$\,K$^2$ and $\alpha_{\rm HI} = -0.6$ for our HI simulation. 

To obtain the simulated patches, we firstly scale  the 2D random Gaussian field with the square root of the power spectrum to introduce scale-dependent structures in our simulation. We then apply the inverse real Fourier transform on the scaled Gaussian field to get the simulated 2D temperature map. The HI covariance matrix for our foreground filter is generated from 10000 realizations of the simulated HI 2D patches  for each frequency channel. We have tested that 10000 realizations are much larger  than the degree of freedom in  the prior covariance matrix to yield unbiased results in our analysis.  The lower left panel in Fig.\,\ref{fig:Tmaps} shows one realization of the simulated HI patches. Compared with the upper left panel, the two completely independent simulation approaches for the test dataset and the covariance matrix give consistent maps in terms of both amplitude and morphology. 

\subsubsection{Synchrotron radiation}
Synchrotron radiation arises from the interaction between energetic charged particles and the Galactic magnetic field \citep{rl79}. At low radio frequencies, synchrotron emission becomes the dominant emission from the sky, brighter than other emissions. The Galactic synchrotron emission has a smooth frequency spectrum that can be approximated by a power-law so that the brightness temperature scales with the frequency as $T\propto\nu^{\beta}$ where $\beta$ is the spectra index varying across the sky \citep{pbm+03}.

We use the reprocessed all-sky \cite{hks+81} 408\,MHz Galactic synchrotron map \citep{rdb+15}  as the template for simulating Synchrotron maps as our test dataset.  The synchrotron map at each frequency is generated by a frequency scaling of the 408\,MHz template map as
\begin{equation}\label{equ:syn}
  T(\nu, \hat{n}) = T_{408\rm MHz}(\nu, \hat{n})\left(\frac{\nu}{408\,\rm MHz}\right)^{\beta(\hat{n})}\,, 
\end{equation}
where the spatially varying spectra index $\beta(\hat{n})$ is estimated from the all-sky spectral index map by \cite{pbm+03}, which has a mean value of $\bar{\beta} = 2.695$ with a standard deviation of $\sigma_{\beta} = 0.120$. For each frequency, we project a 2D $30^{\circ}\times30^{\circ}$ patch at the 10 sky locations described above as our synchrotron test dataset. The upper panel in the second column of Fig.\,\ref{fig:Tmaps} shows one patch of the synchrotron map at the frequency of 400\,MHz. The colorbar highlights the amplitude difference between the synchrotron emission and the HI signal. The foregrounds thus must be properly subtracted in order to detect the HI signal.  

To get the prior covariance matrix of the synchrotron emission, we adopt the simple 2D simulation as described above for HI. The amplitude $A_{\rm syn}$ and scale factor $\alpha_{\rm syn}$ for the simulated synchrotron power spectrum (Equ.\,\ref{equ:simps}) are $5\times10^{-5}$\,K$^2$ and $-5$ respectively. Since the synchrotron maps are correlated across frequency, we generate 10000 realizations of simulated synchrotron maps for the first frequency at 400\,MHz. For each realization, we scale the map to other frequency channels following Equ.\,\ref{equ:syn}, where we replace the 408\,MHz synchrotron map by each realization and scale with respect to 400\,MHz. The spectral index $\beta(\hat{n})$ in this case varies spatially  with a mean of $-2.8$ and a standard deviation of 0.5. The values are chosen so that they are close to the observed values but with larger variations. This is to introduce more complex characters than the test dataset in order for the foreground filter to effectively tackle the complex structures inherent in the synchrotron emission. The lower panel in the second column of Fig.\,\ref{fig:Tmaps} shows one realization of the simulated synchrotron patches at 400\.MHz. The amplitude and large scale diffused structures are consistent with the Haslam-based synchrotron map in the upper panel.

\subsubsection{Free-Free radiation}
Free-free radiation originates from the unbound interaction between free electrons and ions from ionised interstellar medium \citep{rl79}. The free-free frequency spectrum is well-defined by a power-law with an independent spectral index to the Galactic synchrotron emission \citep{ddd03}. Therefore, free-free emission adds spectral curvature to the foreground components, increasing the complexity and difficulties of component separation.

At radio frequencies, the optical $H_{\alpha}$ line is a good tracer of free-free emission at intermediate and high Galactic latitudes ($|b|\gtrsim10^{\circ}$) outside of the Galactic plane. We use the all-sky $H_{\alpha}$ emission map \citep{ddd03} to simulate free-free maps through the $H_{\alpha}$-to-radio relation
\begin{equation}\label{equ:free}
T \approx 10\,\mathrm{mK} \left(\frac{T_e}{10^4\,\rm K}\right)^{0.667}\left(\frac{\nu}{\rm GHz}\right)^{-2.1} I_{H_{\alpha}}\,,
\end{equation}
where $I_{H_{\alpha}}$ is the $H_{\alpha}$ template in Rayleigh and $T_e$ is the electron temperature fixed at 7000\,K for our simulation, which is the typical temperature of warm ionised gas at radio frequencies \cite{add+12}. A 2D $30^{\circ}\times30^{\circ}$ patch is projected at each of the selected 10 sky locations as our free-free test dataset.  The upper panel in the third column of Fig.\,\ref{fig:Tmaps} shows one patch of the free-free maps at 400\,MHz as an example. Compared with the small-scale HI signal and the diffused synchrotron emission, the free-free map has more clustered structures. 

For the covariance matrix of the free-free emission, we follow the 2D simulation as for the synchrotron emission. The amplitude and scale factor based on Equ.\,\ref{equ:simps} are $A_{\rm free} = 10^{-5}$\,K$^2$ and $\alpha_{\rm free} = -2.5$ respectively. The value of the scale factor in this case is between the values of the HI and synchrotron scale factors. This is because free-free emission is less diffused on large scales than the synchrotron emission, but has more clustered structures than the small-scale dominated HI emission. We generate 10000 realizations at the frequency channel of 400\,MHz, and scale to other frequencies with a spatially varying spectral index with a mean of -2.1 and a standard deviation of 0.5. This is consistent with the observed spectral index in Equ.\,\ref{equ:free} but includes more spatial complications to build an effective foreground filter. The lower panel in the third column of Fig.\,\ref{fig:Tmaps} shows one realization of the simulated free-free map at 400\,MHz. We have chosen the spectral index to introduce the medium-scale structures in the simulated map in order to be consistent with the $H_{\alpha}$-based free-free map in the upper panel. 

\subsubsection{Point sources} \label{sec:p_source}
Another component of foreground contamination is the extragalactic point sources consisting of radio galaxies, quasars and other objects. We use the model from \cite{bbd+13} based on observed data from continuum surveys at 1.4\,GHz between 1985 and 2009 to simulate point source maps.  The mean background brightness temperature of the point sources can be modelled by
\begin{equation}\label{equ:Tbg}
\bar{T}_{\rm ps} = \left(\frac{\mathrm{d} B}{\mathrm{d}T}\right)^{-1}\int_0^{S_{\rm max}}S\frac{\mathrm{d}N}{\mathrm{d}S}\mathrm{d}S\,
\end{equation}
where $\mathrm{d} B/\mathrm{d}T = 2\kappa_B\nu^2/c^2$, with $\nu$ being the observing frequency, $c$ being the speed of light and $\kappa_B$ being the Boltzmann constant.  $S_{\rm max} $ is the flux density assuming one can subtract sources with $S > S_{\rm max}$. In principle, one expects to subtract the brightest radio sources down to $S_{\rm max} = 10\,$mJy using the National Radio Astronomy Observatory Very Large Array Sky Survey (NVSS) with a completeness of 3.4 mJy. We choose a conservative value of $S_{\rm max} = 1$\,Jy in our case to test our foreground filter. The source count $\mathrm{d}N/\mathrm{d}S$, quantifying the number of sources per steradian per unit flux, is computed using a 5th order polynomial model from \cite{bbd+13} by fitting observed data of multiple continuum surveys at 1.4\,GHz. 

The fluctuations on the background temperature can be characterised in two components: i) the Poisson distributed sources; ii) the clustered sources.  Poisson distributed sources contribute to the fluctuations in two parts. For weak sources at the limit of a sufficiently large number density, the intensity distribution can be approximated by a Gaussian distribution with a white power spectrum of \cite{bbd+13} 
\begin{equation}\label{equ:clps}
  C_{\ell}^{\rm Poisson} = \left(\frac{\mathrm{d}B}{\mathrm{d}T}\right)^{-2}\int_0^{S_{\rm PS}}S^2\frac{\mathrm{d}N}{\mathrm{d}S}\mathrm{d}S,
\end{equation}
where $S_{\rm PS} = 0.01$\,Jy  is the upper limit of the source flux density that still satisfies a Gaussian distribution \citep{odb+18}. For sources with a higher flux density of  $S_{\rm PS} < S < S_{\rm max}$, the source density becomes too low that we must simulate their contribution by directly distributing sources on the sky map with the number of sources and their flux densities respecting the source count model. To do this, we calculate the number of sources, $N_i$, in aggregate source density bins, $S_i$, between $S_{\rm PS}$  and $S_{\rm max}$ through $N_i = \int_{S_i-\Delta S/2}^ {S_i+\Delta S/2}\frac{\mathrm{d}N}{\mathrm{d}S}\mathrm{d}S$, where $\Delta S$ is the source flux density bin width. For each bin $S_i$, we assign $N_i$ sources with random flux density between $S_i-\Delta S/2$ and $S_i+\Delta S/2$, and distribute them on random sky locations. The corresponding brightness temperature at a particular pixel at location $\hat{n}$ on the sky is computed by
\begin{equation}\label{equ:Tps}
  T_{\rm PS} (\nu,\hat{n}) = \left(\frac{\mathrm{d}B}{\mathrm{d}T}\right)^{-1}\Omega_{\rm pix}^{-1}\sum^J _{j = 1}S_j(\nu),
\end{equation}
where $\Omega_{\rm pix}$ is the pixel size, $J$ is the total number of sources allocated within the pixel, and $S_i(\nu)$ is the flux of each point source at frequency $\nu$.

The power spectrum of clustered point sources can be estimated as \citep{odb+18}
\begin{equation}\label{equ:clcluster}
  C_{\ell}^{\rm Cluster} \approx 1.8 \time 10^{-4}\ell^{-1.2}\bar{T}_{\rm PS}^2\,.
\end{equation}
In summary, the point source map is a combination of a background mean temperature given by Equ.\,\ref{equ:Tbg}, a Gaussian map realization from the power spectrum in Equ.\,\ref{equ:clps} for weak Poisson distributed sources, randomly located strong Poisson sources from Equ.\,\ref{equ:Tps}, and a Gaussian map realization of clustered point source power spectrum in Equ.\ref{equ:clcluster}. 

We adopt a power law to scale the point source brightness temperature into different frequencies by $T_b \propto \nu^{\alpha}$. The spectral index $\alpha$ is randomised for each pixel of the simulated map following a Gaussian distribution
\begin{equation}
  G(\alpha) =\frac{1}{\sqrt{2\pi\sigma^2}}\exp\left[-\frac{(\alpha-\alpha_0)^2}{2\sigma^2}\right]\,,
\end{equation}
where the mean and standard deviation of the Gaussian distribution are $\alpha_0 = -2.7$ and $\sigma = 0.2$ respectively \cite{bbd+13}. As for the other sky components, a 2D $30^{\circ}\times30^{\circ}$ patch is projected at each of the selected 10 sky locations to be the point source test dataset. The upper right panel in Fig.\,\ref{fig:Tmaps} shows one patch of the point source maps at 400\,MHz, which is dominated by small scale structures. 

For the covariance matrix, we simulate 2D point source maps in a much simpler way. We generate a total number of $2\times\rm N_{pix}^2$ point sources, where  $N_{\rm pix} = 150$ is the number of pixels per side. We assume each point source is smaller than the pixel size so that each source occupies a single pixel. We randomly distribute the point sources on the 2D map. Therefore, we have $\sim2$  point sources in each pixel on average. Each source has a randomly allocated brightness temperature between $T_{\rm min} = 0.01$\,K and $T_{\rm min} = 10$\,K. These thresholds are chosen so that the simulated 2D maps have approximately the same order of magnitude as the point source test dataset.  We simulate  10000 realizations of the 2D point source maps at the first frequency channel of 400\,MHz. Each realization is scaled to other frequencies through a spatially varying spectral index with a mean of -2.7 and a standard deviation of 0.5. The lower right panel of Fig.\,\ref{fig:Tmaps} shows one realization of the simulated point source at 400\,MHz. The amplitude is consistent with the test dataset in the upper panel, while the map constitutes of Poisson distributed sources only.

\subsection{Instrument model} \label{instr}

A telescope array object in the simulation pipeline
is characterized by an array layout (the physical
arrangement of the antenna elements) and the parameters
that describe each antenna element in the array including 
primary beam, system temperature, and frequency of operation.
Visibilities are calculated via a two-dimensional
Fourier transform of the flat sky maps weighted by the
telescope's primary beam.

The primary beam is calculated from a user-defined 
window function that represents the 
antenna illumination pattern, and that is Fourier
transformed, interpolated to each pixel in the 
flat-sky map, and squared to obtain the power beam
at each frequency of operation. 
By using an
illumination pattern as a starting point for the
beam, we are able to generate beams that have the desired
frequency dependence and properties of a realistic beam
while having control over the leak of signal power in visibility
space outside the telescope's physical dimensions due to 
the truncated nature of the flat sky maps.
By default, we use a two-dimensional modified Bartlett-Hann window
as the illumination pattern.

The visibilities are corrupted by instrumental noise 
that is modelled as additive complex-valued Gaussian
distributed noise that is stationary and 
uncorrelated between antennas
and frequencies. The noise of each visibility
is determined by the system temperature, integration time, 
bandwidth, and redundancy according to the radiometer equation.

\subsection{Power spectrum estimator}\label{sec:ps_estimator}

In order to quantify the performance of our foreground filter, we compare the power spectra of the recovered HI map and the input HI dataset. The power spectrum estimator is constructed in the form of a quadratic estimator such that \citep[e.g.,][]{dod03}
\begin{equation}\label{equ:ps}
C_{\ell}  =  \frac{F_{\rm \ell\ell'}^{-1}}{2}\left(\mathbf{d_1^{\dagger}}\mathbf{C}^{-1}\mathbf{C_{,\ell'}}\mathbf{C}^{-1}\mathbf{d_2} \right)\,.
\end{equation}
In our case,  $\mathbf{d_1}$ and  $\mathbf{d_2}$ are the recovered HI maps from two different seasons so that the thermal noise will be cancelled out through the cross-spectrum estimation.  $F_{\rm \ell\ell'}$ is the Fisher matrix defined as
\begin{equation}
F_{\ell\ell'}  = \frac{1}{2}\mathrm{Tr}[\mathbf{C_{,\ell}}\mathbf{C}^{-1}\mathbf{C_{,\ell'}}\mathbf{C}^{-1}]\,.
\end{equation}
In each case, $\mathbf{C}^{-1}$ is the inverse covariance matrix including all components ($\mathbf{C} = \mathbf{S} +  \mathbf{F} + \mathbf{N}$), computed using the simplified simulation with 10000 realizations as described in section\,\ref{sec:sky}. Physically, the inverse covariance matrix applies weighting to the data. $\mathbf{C_{,\ell}}$ is the derivative matrix with respect to the HI signal at the angular scale of multipole $\ell$.  Physically, the derivative matrix characterises the properties of the HI signal to enable the accurate estimation of the HI power from the given data. The derivative matrix is calculated by replacing the HI power spectrum in Equ.\,\ref{equ:simps} with a top-hat function so that for each multipole bin centred at $\ell_i$ with a bin width of $\delta\ell$, the power spectrum within that bin is
\begin{equation}
  P_{\ell_i}  = \begin{cases}
    1 & \ell_i-\frac{\delta\ell}{2}<\ell<\ell_i+\frac{\delta\ell}{2} \\
    0 & \rm elsewhere \\ 
    \end{cases}\,.
\end{equation}
Based on our telescope configuration, the largest angular scale we can measure is $\ell = 59$ and the smallest scale is  $\ell = 414$. In our analysis, we choose 14 equally spaced multipole bins within the measurable scales to compute our power spectra. Since our HI simulation is completely Gaussian, the uncertainty on the estimated power spectrum is the square root of the inverse of the Fisher matrix such that
\begin{equation}
  \Delta C_{\ell} = \sqrt{F_{\ell\ell'} ^{-1}}\,.
\end{equation}

Our power spectrum estimator is independent of the exact space of the data, as long as it is consistent with the total covariance and derivative matrices  throughout Equ.\,\ref{equ:ps}. In our analysis, we project the total covariance and derivative matrices into the KL space to estimate the power of the recovered HI signal after our KL-based foreground filter. The projected data and covariance contain all measured degrees of freedom from both baselines and frequencies. Therefore, the estimated power spectrum following Equ.\,\ref{equ:ps} is equivalently the redshift-averaged spectrum of the recovered HI signal.  We compare the recovered HI power spectrum with the input HI spectrum computed from the full-sky HI test dataset simulated with the realistic sky model before projecting into the 2D patches (see section\,\ref{sec:HIsim}). We use the \textsc{anafast} module provided by the \textsc{healpix} package to calculate the input HI power spectrum at the central frequency to be comparable with the recovered HI spectrum.

\bibliography{journals,lit}
\end{document}